\documentclass[sigconf]{acmart}

\usepackage{tikz}
\usepackage{calc}
\usepackage{subfigure}
\usepackage{listings,amsfonts}
\usepackage{amsmath,xcolor,pifont}
\usepackage{url}
\usepackage{balance}
\usepackage{xspace}
\usepackage{hyperref,epsfig,endnotes}
\usepackage{xcolor}
\usepackage{booktabs}
\usepackage{array}
\usepackage{algorithm}
\usepackage{paralist}
\usepackage{multirow,makecell}
\usepackage{enumitem}

\newcommand{\latinlocution}[1]{\textit{#1}}
\newcommand{\eg}{\latinlocution{e.g.,}\xspace}
\newcommand{\ie}{\latinlocution{i.e.,}\xspace}
\newcommand{\etal}{\latinlocution{et al.}\xspace}

\newcommand{\lili}[1]{{\color{red} Li: #1}}
\newcommand{\nv}[1]{{\color{red} Narseo: #1}}
\newcommand{\haoyu}[1]{{\color{blue} HW: #1}}
\newcommand{\jet}[1]{{\color{red} Juan: #1}}

\newcolumntype{L}[1]{>{\raggedright\let\newline\\\arraybackslash\hspace{0pt}}m{#1}}
\newcolumntype{C}[1]{>{\centering\let\newline\\\arraybackslash\hspace{0pt}}m{#1}}
\newcolumntype{R}[1]{>{\raggedleft\let\newline\\\arraybackslash\hspace{0pt}}m{#1}}

\usepackage{adjustbox}
\newcolumntype{R}[2]{%
    >{\adjustbox{angle=#1,lap=\width-(#2)}\bgroup}%
    l%
    <{\egroup}%
}
\newcommand*\rot[2]{\multicolumn{1}{R{#1}{#2}}}

\renewcommand{\texttt}[1]{%
 \begingroup
 \ttfamily
 \begingroup\lccode`~=`/\lowercase{\endgroup\def~}{/\discretionary{}{}{}}%
 \begingroup\lccode`~=`[\lowercase{\endgroup\def~}{[\discretionary{}{}{}}%
 \begingroup\lccode`~=`.\lowercase{\endgroup\def~}{.\discretionary{}{}{}}%
 \catcode`/=\active\catcode`[=\active\catcode`.=\active
 \scantokens{#1\noexpand}%
 \endgroup
 }

\begin{document}

\title[A Large-Scale Comparative Study of Chinese Android App Markets]{Beyond Google Play: A Large-Scale Comparative Study of Chinese Android App Markets}

\author{
Haoyu Wang$^{1}$, Zhe Liu$^{2}$, Jingyue Liang$^{2}$, Narseo Vallina-Rodriguez$^{3,4}$\\Yao Guo$^{2*}$, Li Li$^{5}$, Juan Tapiador$^{6}$, Jingcun Cao$^{7}$, Guoai Xu$^{1}$
}\thanks{*Corresponding author.}

\affiliation{%
  \institution{
$^{1}$ Beijing University of Posts and Telecommunications,\\
$^{2}$ Key Laboratory of High-Confidence Software Technologies (MOE), Peking University,
$^{3}$ IMDEA Networks,
$^{4}$ ICSI,\\
$^{5}$ Monash University,
$^{6}$ Universidad Carlos III de Madrid,
$^{7}$ Indiana University Bloomington
}
}

\renewcommand{\shortauthors}{Haoyu Wang et al.}

\copyrightyear{2018} 
\acmYear{2018} 
\setcopyright{acmcopyright}
\acmConference[IMC '18]{2018 Internet Measurement Conference}{October 31-November 2, 2018}{Boston, MA, USA}
\acmBooktitle{2018 Internet Measurement Conference (IMC '18), October 31-November 2, 2018, Boston, MA, USA}
\acmPrice{15.00}
\acmDOI{10.1145/3278532.3278558}
\acmISBN{978-1-4503-5619-0/18/10}

\begin{abstract}

China is one of the largest Android markets in the world. 
As Chinese users cannot access Google Play to buy and install Android apps, 
a number of independent app stores have emerged and compete in the Chinese app market.
Some of the Chinese app stores 
are pre-installed vendor-specific app markets (\eg Huawei, Xiaomi and OPPO),
whereas others are maintained by large tech companies (\eg Baidu, Qihoo 360 and Tencent). 
The nature of these app stores and the content available 
through them vary greatly, including their trustworthiness and security guarantees. 

As of today, the research community has not studied the Chinese Android ecosystem in 
depth. To fill this gap, we present the first large-scale comparative study 
that covers more than 6 million Android apps downloaded from 16 Chinese app markets 
and Google Play. We focus our study on 
catalog similarity across app stores, their features, publishing
dynamics, and the prevalence of various forms of misbehavior (including
the presence of fake, cloned and malicious apps). Our findings also suggest
heterogeneous developer behavior across app stores, in terms of code
maintenance, use of third-party services, and so forth. 
Overall, Chinese app markets perform
substantially worse when taking active measures to protect mobile users and
legit developers from deceptive and abusive actors, showing 
a significantly higher prevalence of malware, fake, and cloned
apps than Google Play. 
\end{abstract}



\maketitle

\vspace{-0.1in}
\section{Introduction}

According to recent reports, there are more than 
700 million Android users in China~\cite{smartphone}.
Due to the restriction of Google's services in China since late 2010 --
and by extension of Google Play~\cite{block,Censorship} -- 
hundreds of millions of Chinese Android users resort to alternative 
markets to buy and install Android apps. 
This restriction over Google services in China has been seen as 
a business opportunity by many Chinese 
Internet companies (\eg Tencent and Baidu) and smartphone manufacturers (\eg 
Huawei and Xiaomi).
Despite the fact that these app markets target mainly the Chinese Android users, they are also available to users from all over the world. 

The diversity and large number 
of third-party markets in China have made it difficult
for both mobile users and app developers to choose 
the most appropriate one(s) to discover or distribute their
apps. 
This state of affairs has also
opened new opportunities for malicious actors:  
previous work has suggested that repackaged apps, including malware, 
are widely distributed in Google Play, but especially through
third-party markets~\cite{RiskRanker, DroidMoss, Juxtapp, Dong-FSE-18, AdDarwin, wukong, hu2018dating}.

To the best of our knowledge, no previous work has
performed a systematic and 
comparative study across different app markets, including the Chinese ones. 
To fill this research gap, we perform a multi-dimensional and 
large-scale study covering more than 6.2 million apps to identify the 
differences between Google Play and 16 popular Chinese Android app markets. 
We begin our study by offering a
high-level characterization of these app stores, discussing features such as
their copyright checks, app auditing processes, 
their strategies to attract app developers, 
and their transparency efforts (Section~\ref{sec:features}). 
Second, after presenting our dataset and  
app collection method (Section~\ref{sec:crawling}),
we compare their download distributions, 
user rating distributions, and presence of third-party tracking and advertising libraries (Section~\ref{sec:overview}).
Third, we study their catalog similarities and their
publication dynamics, 
with emphasis on detecting the distribution of the presence of a given 
developer and app across stores (Section~\ref{sec:publishing}).  
We then provide an in-depth analysis of malicious and deceptive
behaviors across app markets, discussing the presence of fake 
and cloned apps, over-privileged apps, and malware (Section~\ref{sec:misbehaviors}).
We conclude our paper with a discussion around the state
of affairs in the Chinese Android ecosystem, and its
implications to users and developers alike (Section~\ref{sec:post}). 

Our main research contributions are as follows:
\vspace{-0.05in}

\begin{itemize}[leftmargin=*]
\item We conduct a comparative study of various intra- and inter-market features.
Our results reveal a long tail distribution of app
popularity, with the top 1\% of apps usually accounting for over 80\% of total downloads
across the 17 studied markets. Further, we observe the presence of
heterogeneous behaviors across markets (\eg 
in terms of code maintenance and metadata consistency). 

\item We find that the set of third-party libraries 
(\eg SDKs provided by advertising and tracking services)
embedded in Android apps 
are different for those published in Chinese stores than in Google Play. 
This observation could be explained by the inability to access 
Google services such as Google Analytics and AdMob
from China, and Chinese developers' need to monetize their apps through 
services specialized in the Chinese market.   

\item Popular apps
are more likely to be simultaneously published in multiple markets compared to unpopular
ones. However, there is a strong market bias across developers:
1) 57\% of Google Play developers do not publish their 
apps on any of the Chinese markets, and 2) almost half of the Chinese-specific 
developers do not publish apps in Google Play. 

\item We analyze the prevalence of various types of malicious behaviors in our dataset,
specifically the presence of fake and cloned apps, over-privileged apps, and malware.
Google Play clearly outperforms Chinese markets in all dimensions of our study thanks to
their positive efforts to eradicate these behaviors. Our results reveal
that the presence of malicious and repackaged apps in the majority of
Chinese app stores is significant and prevalent over time 
(10\%, on average, in the case of malware), in some cases reaching almost 1 in 4 apps in the market. 

\item In order to estimate the extent to which app markets implement security checks on submissions, we performed a second crawl 8 months after the first snapshot. Our exploration suggests that
over 84\% of the potentially malicious apps found in Google Play were removed. This differs
considerably in the case of Chinese markets, with malware removal ratios varying from 0.01\%
to 34.51\% in the best case.

\end{itemize}

To the best of our knowledge, this is the first comparative
study between Google Play and alternative Chinese app stores at scale,
longitudinally and across various dimensions. 
Our results motivate the need for
more research efforts to illuminate the widely
unexplored Chinese mobile and web ecosystem. 
We believe that our efforts 
can positively contribute to bring user and developer awareness, 
attract the focus of the research community and regulators, and
promote best operational practices across app store operators.
We have released our dataset, along with the experiment results, 
to the research community at:
\textbf{\url{http://market.orangeapk.com/}}

\section{Chinese Android App Markets}
\label{sec:features}

Due to the access restrictions of Google Play in China, Chinese Android users 
resort to a large ecosystem of 
alternative third-party Android app markets, which could be
classified into three categories according to their nature:

\begin{itemize}[leftmargin=*]
  \item \emph{\textbf{Vendor-specific app markets.}} 
  China has a vast and powerful smartphone manufacturing industry with well-known vendors such as 
  Huawei, Xiaomi, and Lenovo. 
  Almost every Chinese smartphone vendor maintains its own 
  app market, which also comes pre-installed on their devices.
  \item \textbf{\emph{Web companies.}} 
  Chinese Internet giants such as 360, Baidu, and Tencent also compete 
  in the Chinese Android market with their own app stores. 
  These companies usually provide support to some smartphone vendors behind
  the scenes. For example, the Sony app store in China is powered by Baidu App Market, 
  and the Smartisan app store is supported by Tencent Myapp Market. 
  \item \textbf{\emph{Specialized markets.}}
  A number of relatively small Chinese companies are specialized in 
  Android app services. They usually make profit through app promotion and other business-oriented partnerships with app developers/companies.
  For example, 25PP is an Android app market powered by the PP smartphone assistant, 
  which is a popular management system app in China. Similarly,
  Wandoujia is an app store provided by a company
  focused on app recommendation, especially for mobile games. 
\end{itemize}

In this study, we first resort to several independent 
industry reports about app market ranking in China~\cite{MarketReport5, MarketReport6, MarketReport1, MarketReport2, MarketReport3, MarketReport4}.
We cover all the top 10 Android markets in China, excluding the Vivo market (ranks 6 to 10 in China), because the Vivo market does not provide a web-based app download interface, which makes it difficult for us to crawl the apps. 
Our list covers the app stores for the top five smartphone vendors in China~\cite{smartphonemarket}, 
three top Chinese web companies, and eight popular specialized Android app markets. The app markets in this list cover more than 98\% of active users in China~\cite{MarketReport5, MarketReport6}.
The majority of these markets target Chinese Android users, but
some of them operate at a global scale, particularly those run by Android handset vendors. 
For example, Huawei's app market is also popular in Europe, Latin America, and the Middle East~\cite{HuaweiMarket}. 

\subsection{Features of Chinese App Markets}

In this section, we study some critical 
aspects and features across app stores, including their openness to developers, their publication and app auditing process, and their transparency, as shown in Table~\ref{table:marketfeature}. 
For that, we first registered a developer account for each market and 
then manually examined their \emph{developer policies}, \emph{terms of service} and other documents released by these markets~\cite{gplaydeveloper, tencentdeveloper, 
baidudeveloper, 360developer, OPPOdeveloper, Xiaomideveloper, Meizudeveloper, 
Huaweideveloper, Lenovodeveloper, Alideveloper, Anzhideveloper, Liqudeveloper, 
Sogoudeveloper, AppChinadeveloper}.

\begin{table*}[t!]
\small
\newcommand{\tabincell}[2]{\begin{tabular}{@{}#1@{}}#2\end{tabular}}
\caption{Dataset size and market features for Google Play and the 16 Chinese markets studied in this paper.}

\centering
\resizebox{\linewidth}{!}{

\begin{tabular}{l l r r r r c c c c l c c c c c c c c}

\textbf{Market} &
\textbf{Type} &
\rot{90}{1em}{\textbf{Size (\#Apps)}} &
\rot{90}{1em}{\textbf{
\tabincell{l}{Aggregated\\Downloads}
}} &
\rot{90}{1em}{\textbf{
\tabincell{c}{\#Developers}
}} &
\rot{90}{1em}{\textbf{
\tabincell{l}{\% Unique\\Developers}
}} &
\rot{90}{1em}{\textbf{Openness}} &
\rot{90}{1em}{\textbf{
\tabincell{l}{Copyright\\Check}
}} &
\rot{90}{1em}{\textbf{
\tabincell{l}{App\\Vetting}
}} &
\rot{90}{1em}{\textbf{
\tabincell{l}{Security\\Check}
}} &
\rot{90}{1em}{\textbf{
\tabincell{l}{Vetting\\Time}
}} &
\rot{90}{1em}{\textbf{
\tabincell{l}{Quality Rating}
}} &
\rot{90}{1em}{\textbf{Incentive\#1}} &
\rot{90}{1em}{\textbf{Incentive\#2}} &
\rot{90}{1em}{\textbf{Incentive\#3}} &
\rot{90}{1em}{\textbf{
\tabincell{l}{Privacy\\Policy}}
}
&
\rot{90}{1em}{\textbf{
\tabincell{l}{Advertisement}
}}
&
\rot{90}{1em}{\textbf{
\tabincell{l}{In-app\\Purchase}
}}

\\

\midrule

Google Play &
Official &
2,031,946 &
193 B &
538,283 &
57.04 &
\checkmark &
\checkmark &
\checkmark &
\checkmark & 
Hours &
\checkmark &   
 &
\checkmark &
\checkmark &
\checkmark &
\checkmark &
\checkmark \\

\hline

Tencent Myapp & 
Web Co. &
636,265 & 
82 B & 
294,950 & 
10.61 &
\checkmark  & 
\checkmark &
\checkmark & 
\checkmark & 
1 day &
\checkmark & 
\checkmark & 
\checkmark & 
\checkmark & 
& \checkmark 
&
\\ 

Baidu Market & 
Web Co. &
227,454 & 
94 B &
107,698 & 
15.10 &
\checkmark & 
\checkmark & 
\checkmark & 
\checkmark & 
1-3 days & 
&
\checkmark  
& 
& 
& 
& \checkmark &
\\  

360 Market & 
Web Co. &
163,121 & 
50 B & 
90,226 & 
6.80 &
\checkmark &  
\checkmark & 
\checkmark & 
\checkmark &  
1 day &
\checkmark &
\checkmark &
\checkmark & 
& 
& \checkmark &\checkmark
\\

\hline

OPPO Market &
HW Vendor &
426,419 & 
57 B & 
209,197 & 
14.37 &
Partial\footnote{It only allows publishing apps for specific categories.} &
\checkmark & 
\checkmark & 
\checkmark &
1-3 days & 
&
\checkmark & 
& 
& 
& \checkmark
\\ 

Xiaomi Market &
HW Vendor &
91,190 & 
- &
55,669 & 
5.78 &
\checkmark & 
\checkmark & 
\checkmark & 
\checkmark &
1-3 days & 
&
\checkmark & 
& 
&
\\ 

MeiZu Market &
HW Vendor &
80,573 & 
19 B & 
50,451 & 
0.58 &
\checkmark & 
\checkmark & 
\checkmark & 
\checkmark &
1-3 days & 
&
\checkmark & 
& 
& 
\\

Huawei Market &
HW Vendor &
51,303 & 
83 B & 
32,927 & 
5.66 &
\checkmark & 
\checkmark & 
\checkmark & 
\checkmark &
3-5 days & 
&
\checkmark & 
\checkmark & 
\checkmark & 
& \checkmark
\\ 

Lenovo MM &
HW Vendor &
37,716 & 
24 B & 
24,565 & 
0.79 &
 & 
\checkmark & 
\checkmark & 
\checkmark &
2 days & 
&
\checkmark & 
& 
& 
\\ 

\hline 

25PP &
Specialized &
1,013,208 & 
56 B & 
470,073 & 
19.06 &
\checkmark & 
\checkmark & 
\checkmark & 
\checkmark &
1-3 days & 
&
\checkmark & 
\checkmark & 
& 
& \checkmark
\\ 

Wandoujia &
Specialized &
554,138 & 
38 B & 
291,114 & 
0.97 &
\checkmark & 
\checkmark & 
\checkmark & 
\checkmark &
1-3 days & 
&
\checkmark &
\checkmark &
&
\\ 

HiApk & 
Specialized &
246,023 & 
17 B & 
115,191 & 
3.65 &
\checkmark &
N/A &
N/A &
N/A &
N/A &
&
&
&
&
\\

AnZhi & 
Specialized &
223,043 & 
12 B & 
74,145 & 
21.93 &
\checkmark &
\checkmark &
\checkmark &
\checkmark &
1-3 days &
&\checkmark &
&
&
\\ 

LIQU & 
Specialized &
179,147 & 
26 B & 
101,336 & 
6.10 &
\checkmark &
\checkmark &
\checkmark &
\checkmark &
N/A &
&
\checkmark &
&
&
\\

PC Online &
Specialized &
134,863 & 
0.2 B & 
65,225 & 
2.58 &
\checkmark &
N/A &
N/A &
N/A &
N/A &
&
& 
& 
&
\\ 

Sougou & 
Specialized &
128,403 & 
3 B & 
66,759 & 
4.04 &
\checkmark & 
\checkmark & 
\checkmark & 
\checkmark &
1 day & 
& 
\checkmark & 
\checkmark & 
& 
& \checkmark
\\ 

App China & 
Specialized &
42,435 & 
- & 
23,699 & 
3.22 &
\checkmark & 
\checkmark & 
\checkmark & 
\checkmark &
1-3 days & 
& 
& 
& 
& 
& \checkmark
\\

\midrule

\textbf{Total} &
&
6,267,247 & 
754 B & 
1,035,992 & 
&
& 
& 
& 
&
& 
&
& 
& 
& 
\\
\hline 
\end{tabular}
}

\vspace{-0.1in}
\label{table:marketfeature}
\end{table*}

\begin{enumerate}[leftmargin=*]
\item \textbf{Openness: } Most Chinese app markets allow third-party developers
to publish their apps for free\footnote{Google Play's registration fee costs \$25~\cite{GooglePlay25Dollar}.}.
However, a small number of app stores enforce some limitations. 
For instance, Lenovo's MM market only allows registered companies to release apps~\cite{Lenovodeveloper}, 
whereas OPPO market only allows publishing apps
falling in specific categories, such as ``wallpaper'' and ``theme'' apps~\cite{OPPOdeveloper}. 
Vendor markets such as OPPO and Xiaomi force developers to release apps that are fully compatible 
with their own devices~\cite{OPPOdeveloper, Xiaomideveloper}.
Finally, App China explicitly limits an APK size to 50 MB~\cite{AppChinadeveloper}.
 
\item \textbf{Copyright checks: } In order to limit the publication of fake and cloned apps, 
all the Chinese markets but HiApk 
and PC Online perform copyright ownership checks. Developers should submit a 
``Software Copyright Certificate'' indicating that they are the original authors of the released apps. 

\item \textbf{Publishing incentives: } 
Chinese app stores provide a number of incentive mechanisms 
for encouraging app developers to publish their apps. These models could be classified into three categories. 
The first one is \textbf{``The Starting App and Exclusive App Free Promotion''}, a common mechanism 
across markets which gives stores publication exclusivity for a period of time in exchange
for actively taking measures to promote the app, typically during 24 hours.
The second category is \textbf{``High Quality App Free Promotion''}. 
Some markets have a qualification of high-quality apps. 
Apps that meet the criteria to obtain such a qualification could request the markets
to promote them for free. 
The third category is \textbf{``Editors' Choice''}, in which the store 
recommends apps based on personal opinions. 

\item \textbf{Auditing process:} All app markets but HiApk and PC Online 
indicate that apps are published after an inspection and vetting process. 
Moreover, eight markets (Google Play, Tencent, OPPO, Xiaomi, 
Meizu, Huawei, Anzhi and AppChina) claim to incorporate \emph{human inspections} attempting to complement the automated auditing process. The general approach is to use automated security analysis tools first to identify possible threats\footnote{Some web companies have released their own security analysis tools for Android apps, e.g., 360 Mobile Security~\cite{360Security} and Baidu PhoneGuard~\cite{li2017fbs,li2016exploring}.}, and then manually check the most suspicious submissions.
For example, a majority of the top apps in the Huawei market are labeled with a sign indicating that they went through manual inspection before being made publicly available, and it is reported that Huawei has a large human inspection team~\cite{huaweiinspection}. 
Excluding HiApk and PC Online, Chinese alternative app stores also explicitly check and
report security issues on the apps (\eg malware and aggressive adware).
The inspection time varies across markets, from several hours (Google Play) to roughly 5 days (Huawei market). 
360 market requires all the developers to use their packaging tool 360 Jiagubao~\cite{360jiagu} to obfuscate apps before entering the market. 

\item \textbf{App quality ratings:} Only Tencent Myapp market and 360 market explicitly 
report that they rate the quality of published apps based on downloads, 
user comments, developer level and other metrics. 
For high quality apps, they could provide more market resources
(\eg advertise them on the starting page) for app promotion to attract high quality developers.

\item \textbf{Transparency:} As opposed to Google Play, 
none of the Chinese app markets require developers to publish their 
\emph{privacy policies} whenever they obtain and use sensitive user data. 
However, nine markets (Google Play, Tencent Myapp, Baidu, 
360, OPPO, Huawei, 25PP, Sougou and AppChina) explicitly inform users 
whether the apps contain \emph{advertisements}. Only Google Play and 360 market report
the presence of \emph{in-app purchases} in the apps.
\end{enumerate}

\section{APK Collection}
\label{sec:crawling}

We implemented a crawler to harvest APKs from Google Play and the
16 alternative Chinese Android 
app stores listed in Table~\ref{table:marketfeature} in August 2017. 
For each app, we also collect publicly available metadata as 
provided by the app stores, including, among others,
the app name, version name, app category, description, downloads, ratings 
and release/update date.

We follow different strategies to crawl each market. In the case of Google Play, 
we use a list of 1.5 million package names provided by 
PrivacyGrade~\cite{privacygrade} as the \emph{searching seeds}. We
use a breadth-first-search (BFS) approach to crawl (1) additional 
related apps recommended for each one of our seeds by Google Play, and (2) 
other apps released by the same developer. In order
to avoid potential regional bias, 
we instrumented our crawler to support both English and Chinese languages. 
Chinese app markets index apps in different ways. Consequently, 
we adapt our crawler to the indexing behavior of each Chinese app market.
For instance, as of this writing, Baidu's app market indexes apps incrementally\footnote{
We use the following syntax: \url{http://shouji.baidu.com/software/INTEGER.html}. }.

We launched several crawlers in parallel via 50 Aliyun Cloud Servers~\cite{aliyun}
between August 15 and August 30, 2017. 
However, published Android apps can be updated by the developers at any time, potentially
affecting our analysis. To overcome this challenge, we implement a ``\textit{parallel search}'' 
strategy in our crawler. As long as we identify a new app (based on its package name) in one of
the 17 markets, we immediately search this app (using either the app name or its package name, according to different markets) in all the remaining markets to crawl it simultaneously
if found. 
Note that we will crawl all the listed searching results and add them to the searching seeds.
After roughly 8 months, we launched a second, one-week crawling campaign in April 30, 2018
for analyzing whether any of the studied malicious apps 
has been removed from each individual stores (Section~\ref{sec:post}).

\subsection{Dataset}
\label{sec:datasetdescription}

Table~\ref{table:marketfeature} reports the number of harvested APKs 
per store. 
We crawled metadata for 6,267,247 different apps across all app stores, 
and 4,522,411 APK files. 
To the best of our knowledge, our dataset is the largest cross-store
APK collection obtained by the research community.
The mismatch between app metadata and APKs is due to
Google Play's rate limiting mechanism, which limited
our APK collection efforts to a random sample of 
287,110 of them. 
We resorted to AndroZoo~\cite{li2017androzoo, androzoo} to obtain offline the APK files 
for 1,553,382 of the missing Google Play apps, 
using the package name and version name as primary key. 
Note that, although this dataset does not cover all the available apps in these markets
due to the limitation of our BFS app crawling method\footnote{As of this writing, the number of apps in Google Play is 2,893,556~\cite{appbrain}, while we only crawled roughly 70\% of them.}, we believe that our dataset has covered the most
popular apps in both Google Play and the Chinese markets. Further, due to the parallel search strategy, the apps studied across markets will not bias the results.

\section{General overview}
\label{sec:overview}

We now study high-level characteristics of Google Play
and the 16 Chinese app stores. Using Google Play as a reference, we 
briefly discuss differences along various dimensions such as catalog 
diversity, user downloads, Android API support, third-party libraries and
user ratings. 

\subsection{App Categories}

\begin{figure}[t!]
\small
  \centering
  \includegraphics[width=0.95\linewidth]{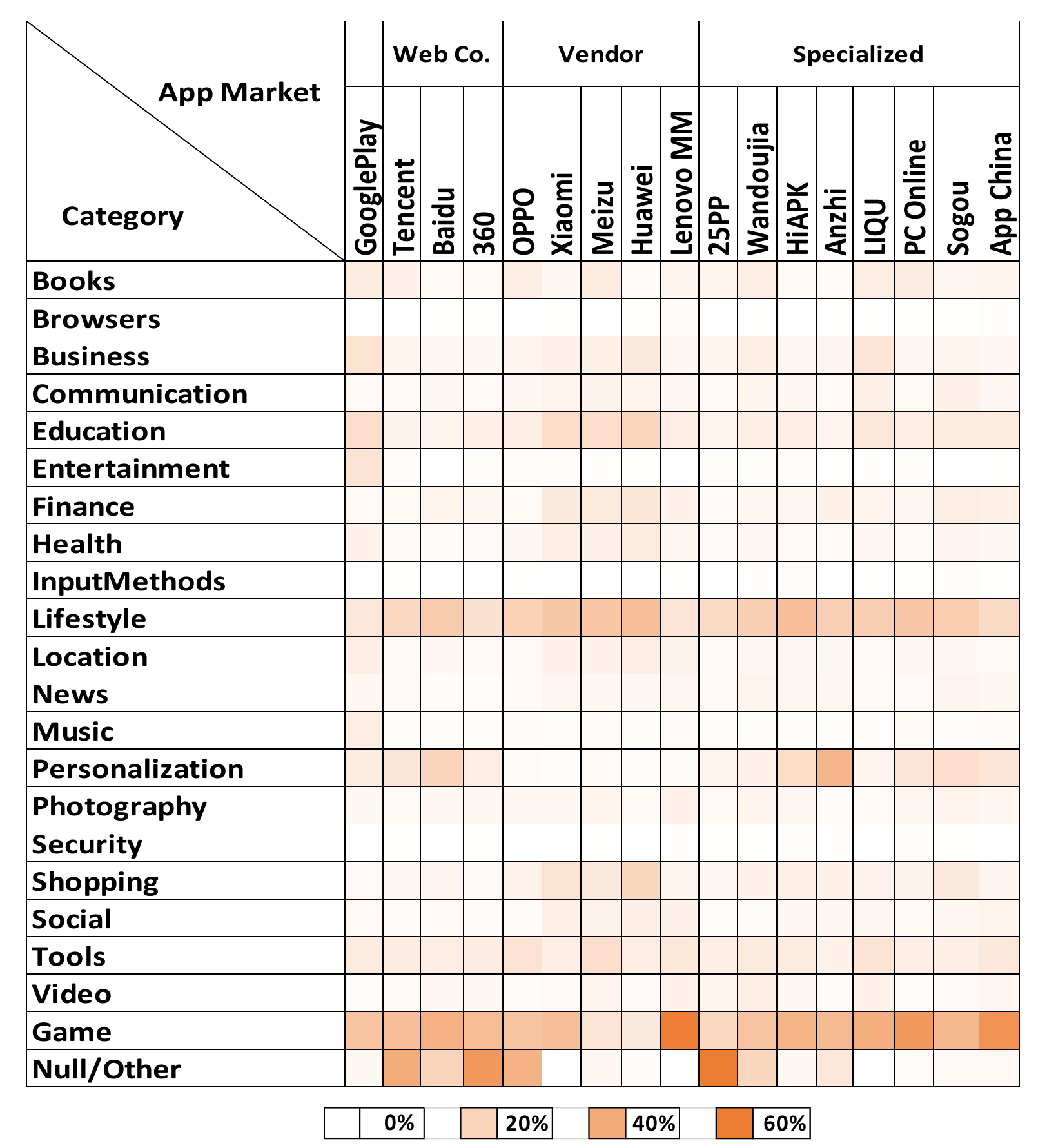}
  \vspace{-0.05in}
  \caption{Distribution of app categories.}
    \vspace{-0.1in}
  \label{fig:category}
\end{figure}

App stores give app developers the freedom to publish their apps
in specific app categories. However, 
each Android market implements a different taxonomy of apps.  
While Google Play defines 33 app categories 
(excluding game app subcategories), 
Huawei Market only has 18 categories. 
In order to perform a fair comparison across markets, 
we manually develop a consolidated taxonomy containing 22 app categories, as shown in Figure~\ref{fig:category}.
Due to the lack of enforcement and lax supervision over the metadata provided 
by app developers, 
in Tencent, 360, OPPO, and 25PP markets, 
we classify 40\% of the apps from these stores as ``Other'' category\footnote{Apps published in these markets can report NULL or non-descriptive categories (\eg 
``Unclassified'', ``102229'').}.

It is noticeable that \texttt{Games} account for roughly 50\% of all apps across markets, while other popular categories include \texttt{Lifestyle} and \texttt{Personalization}. The least popular categories are \texttt{Browsers}, \texttt{InputMethods} and \texttt{Security} tools.
Note, also, how the distribution of published
apps across categories for the majority of Chinese app stores 
follows closely Google Play's distribution. 
A number of app stores, especially vendor ones such 
as Meizu, Huawei and Lenovo's, present a different distribution
of categories.

\subsection{User Downloads}\label{Sec:UserDownloads}

The majority of app stores report the 
actual number of user installs per app while
Google Play bins them into installation ranges (\eg ``50,000 - 100,000'').  
However, this metadata may not be consistent across stores. 
Xiaomi and AppChina do not report this information at all. 
Further, 
we suspect that some of the app stores might be reporting the number of user downloads, likely
higher than the number of user installs, instead of user installs.
For comparison purposes and minimize bias, we normalize
the number of user installed apps for each app store (excluding Xiaomi and AppChina) to Google Play's ranges\footnote{\eg 75,123 after normalization becomes [50,000, 100,000].}.

\begin{figure}[t]
  \centering
  \includegraphics[width=3.4in]{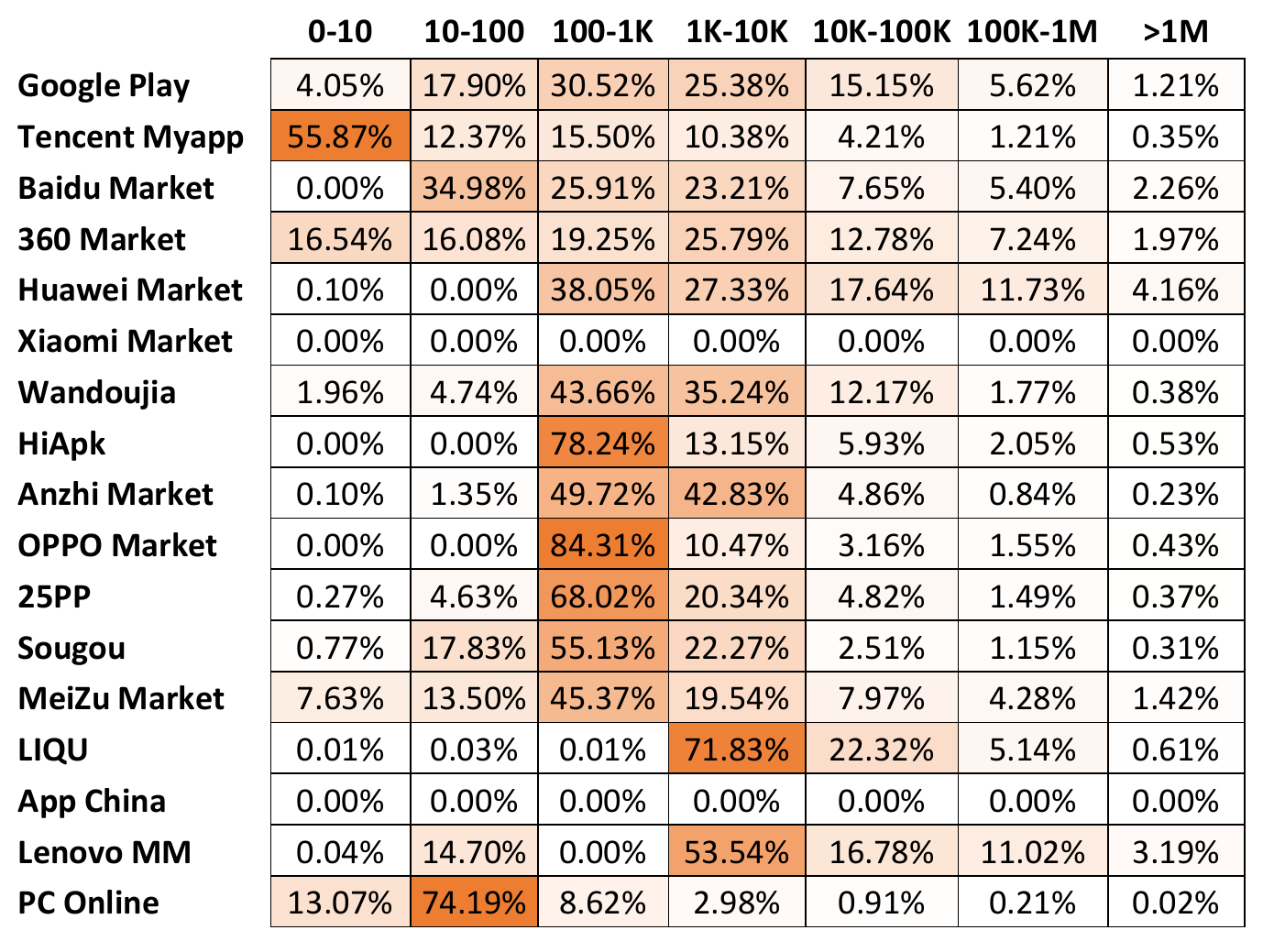}
  \vspace{-0.15in}
  \caption{Distribution of downloads across markets.}
  \vspace{-0.1in}
  \label{fig:downloadsDistribution}
\end{figure}

As Table~\ref{table:marketfeature} reports, 
the apps in Google Play have 193 Billion aggregated downloads\footnote{Estimated by considering the lower bound limit of Google Play's install range.}.  
No Chinese app store gets closer individually to this volume 
despite the size of the Chinese market in terms of user-base. 
However, the number of aggregated downloads across all the 16 studied markets is
three times higher than that of Google Play.  
This figure shows the importance of Chinese Android markets when
aggregated.

The distribution of app downloads follows a power-law distribution,
regardless of the app market, 
as shown in Figure~\ref{fig:downloadsDistribution}.
In general, 85\% of the analyzed apps have less than 10K installs. 
However, subtle differences arise when
looking in detail on a per-store basis after ranking the apps by their number of installs. 
On average, the top 0.1\% of the apps
account for more than 50\% of the total downloads, 
regardless of the app store. However, 
the top 0.1\% of apps published in Tencent MyApp 
account for more than 80\% of the total downloads while
more than 55\% of its published apps have almost no downloads ($<10$).
On the other hand, 15\% of the apps published in
vendor app stores like Huawei's and Lenovo's have more than 100K installs. 
This suggests that there are significant differences in the popularity and
quality of the apps published in certain app stores, as we will investigate later. 

\subsection{Minimum API Level}

\begin{figure}[t]
  \centering
  \includegraphics[width=3.6in]{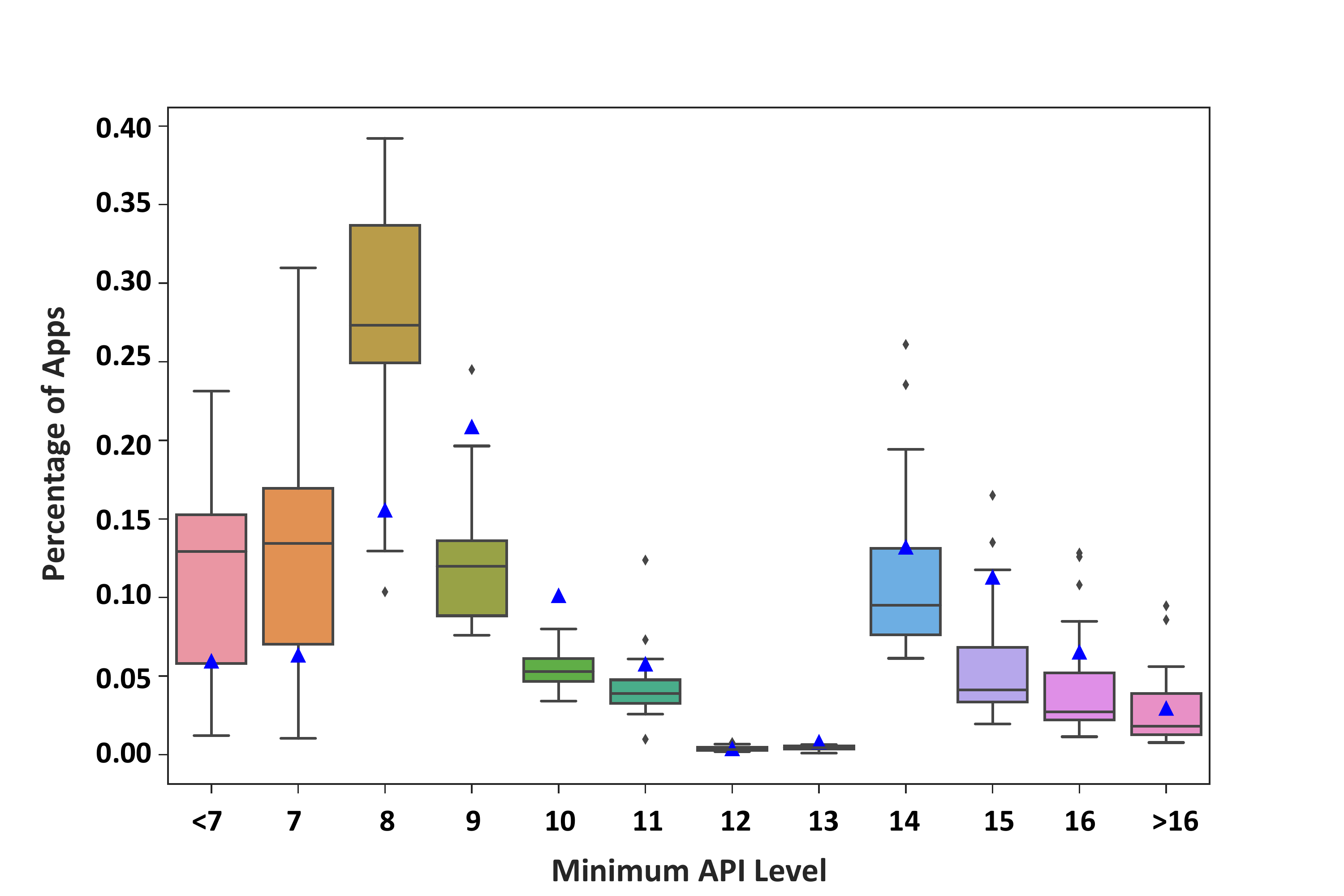}
  \vspace{-0.1in}
  \caption{Distribution of minimum API level declared for each app for the analyzed markets. The triangle symbol represents the value for Google Play, while the box-plots represent the values across the 16 Chinese alternative stores. 
  }
  \label{fig:minAPI}
\end{figure}

Android app developers can declare in the app 
manifest the minimum Android API 
level supported by their apps. This information could offer insights 
about whether app developers are trying to maximize app customers, or whether they 
try to target top-end users.
Figure~\ref{fig:minAPI} shows the distribution of minimum API level declared for each app in each market. The result suggests that API levels 7-9 (\ie Android versions 2.1.x to 2.3.2)
are the most widely supported minimum API levels by the analyzed apps. 
However, the percentage of apps in alternative Chinese markets 
supporting low API levels is 3x higher than that of Google Play in general:
roughly 63\% of apps in Chinese third-party markets support 
API levels lower than 9, as opposed to 22\% in the case of Google Play.

\begin{figure}[t]
  \centering
  \includegraphics[width=3.4in]{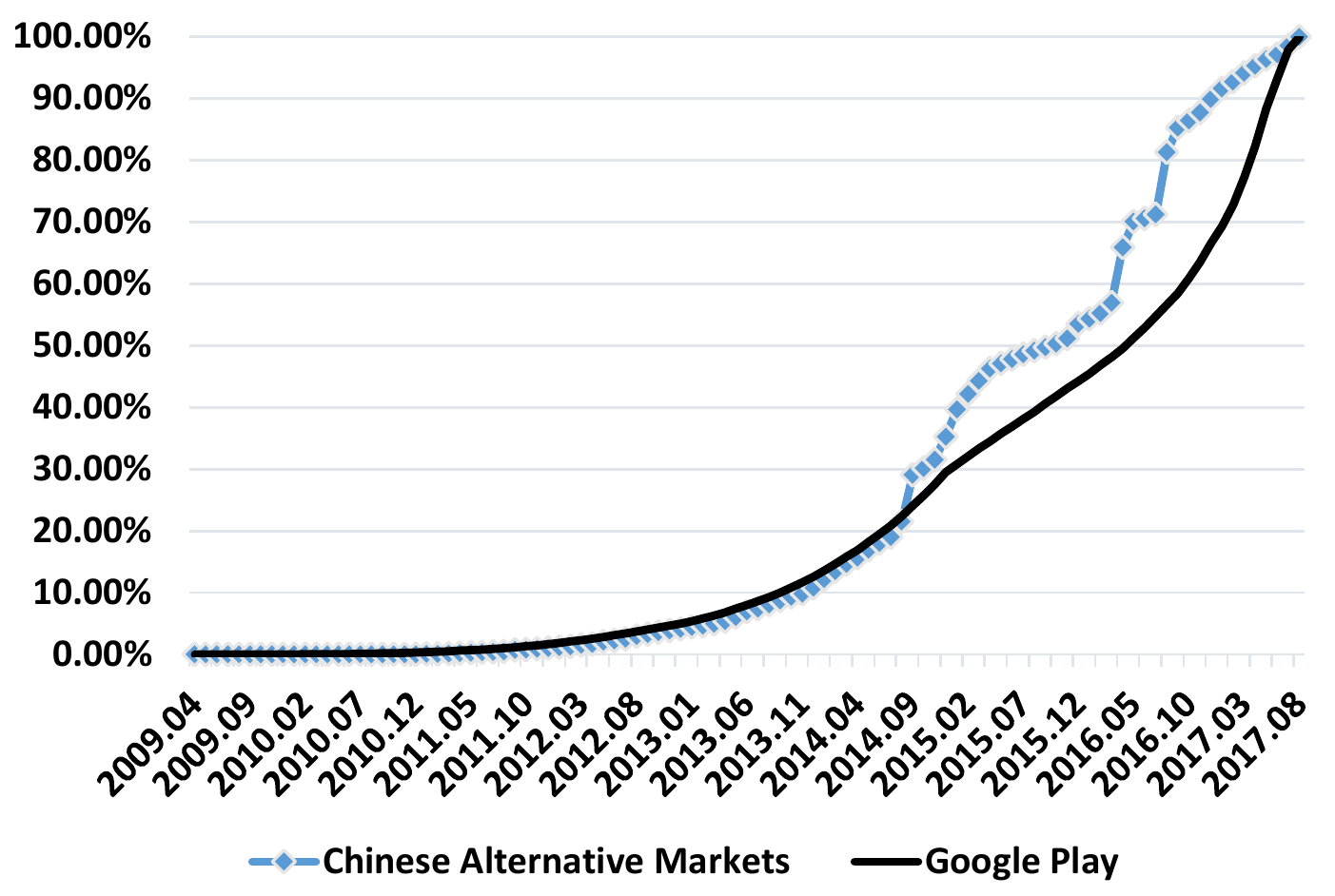}
  \caption{Distribution of app release/update dates. 
  }
  \vspace{-0.1in}
  \label{fig:updateTime}
\end{figure}

We further analyzed the release or update time of these apps 
across markets. This is also a metric used for 
estimating whether developers actively maintain their apps, a strong signal
for code quality~\cite{hassan2017empirical,li2018cid}.
Figure~\ref{fig:updateTime} shows the distribution of
the release/update time of the apps in our dataset, 
as reported by the markets. 
As we can see, roughly 90\% of apps in 
Chinese alternative markets were released/updated 
before 2017, while the number in Google Play is 66\%. 
Further, only 5\% of apps published in Chinese stores
were updated/released within 6 months 
before launching our crawling campaign, 
while more than 23\% of Google Play apps where released during the same time frame. This finding suggests that most of the apps published in Chinese markets likely support
low API-level as they were released years ago. These apps do not 
get timely updates, hence likely exposing
their user-base to various security risks~\cite{hassan2017empirical}, 
and do not take advantage of features introduced in newest Android versions.

\begin{figure*}
\centering
\subfigure[Third-party Libraries]{\label{fig:tpl}
\includegraphics[width=1\columnwidth]{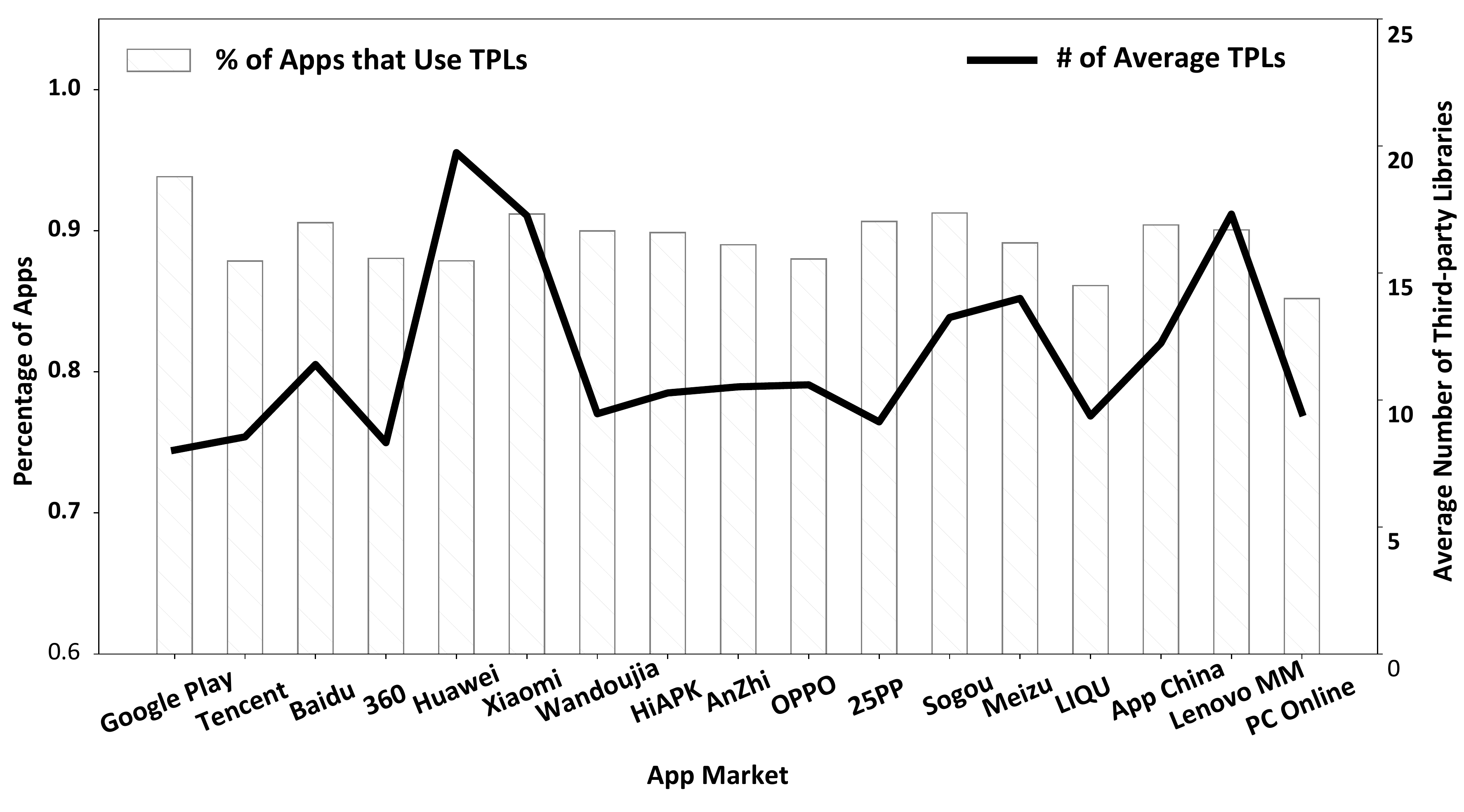}
}
\subfigure[Advertisement Libraries.]{\label{fig:ad}
\includegraphics[width=1\columnwidth]{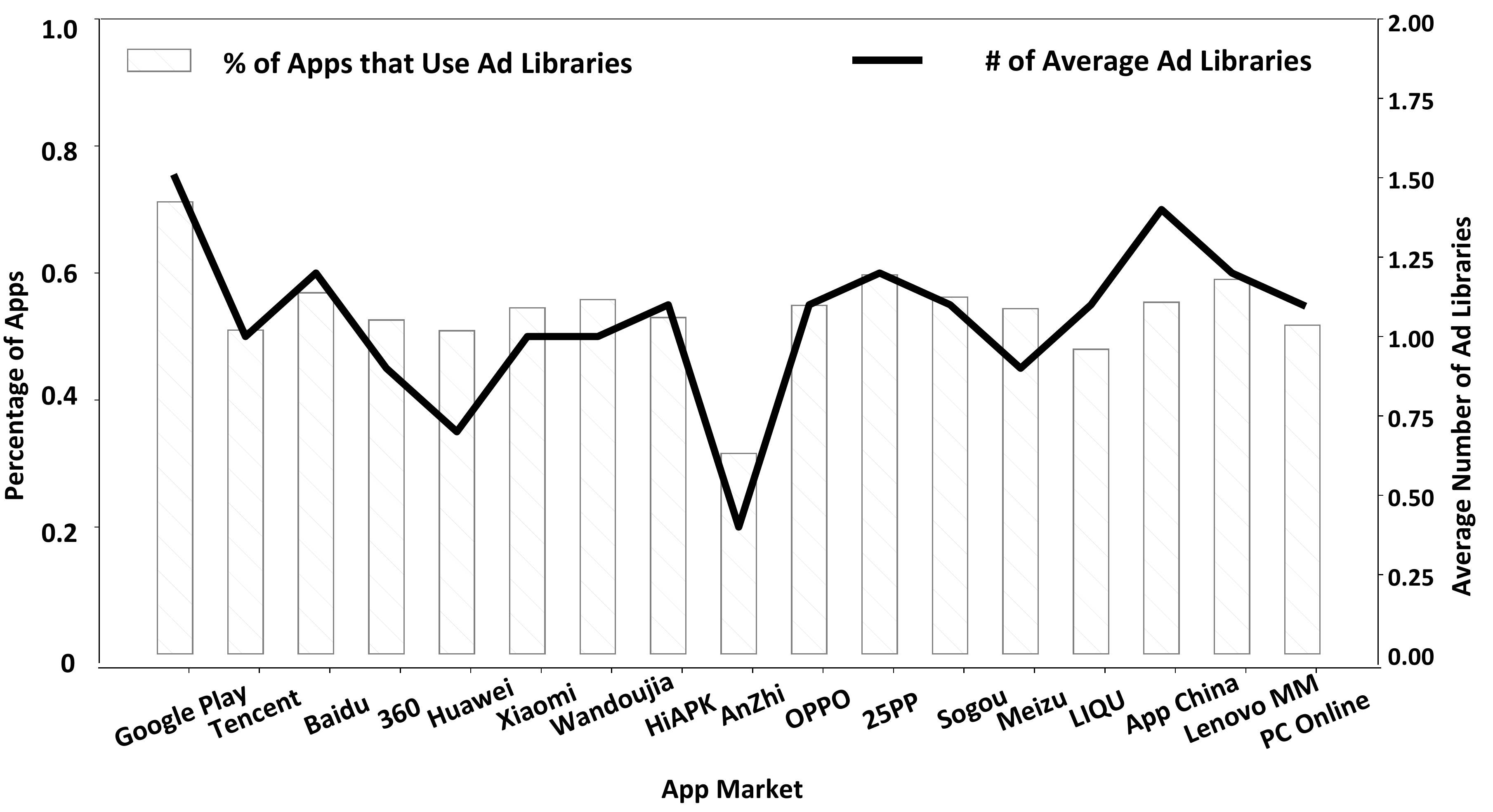}
}
\vspace{-0.1in}
\caption{Presence of third-party libraries across app stores.}
\label{fig:library}
\end{figure*}

\subsection{Third-party Libraries}

Third-party services form an integral part of the
mobile ecosystem: they ease app development and enable features such as analytics, social network integration, and app monetization through advertisements~\cite{vallina2012breaking,razaghpanah2018apps, wang2017understanding}. 
However, aided by the general opacity
of mobile systems, such services are also largely invisible to users, hence
causing potential privacy risks 
~\cite{ReliableLibrary,razaghpanah2018apps,ikram2016analysis,ren2018bug,ren2016recon,wang2017understanding,liu2016identifying}. This is aggravated by the lack of 
transparency enforcement across alternative app stores (Section~\ref{sec:features}): 
no alternative Chinese store requires 
developers to publish a privacy policy, and
only a handful of them actively report the 
presence of ad services or in-app purchases in published apps.

\begin{table}[t]
\small
\centering
\caption{Top 10 third-party libraries for Google Play and Chinese markets apps. Chinese market specific libraries are highlighted in the table.}
\begin{tabular}{|l|l|l|}
\hline
\multicolumn{3}{|c|}{\textbf{Google Play}}                      \\ \hline
\textbf{Package Name  }         & \textbf{Type }              & \textbf{Usage (\%)}  \\ \hline
com.google.android.gms         & Development       & 66.1  \\ \hline
com.google.ads & Advertisement         & 62.1  \\ \hline
com.facebook           & Social Networking   & 21.5   \\ \hline
org.apache        & Development         & 20.5  \\ \hline
com.squareup            & Payment         & 13.8  \\ \hline
com.google.gson & Development & 12.9 \\ \hline
com.android.vending & Payment & 12.5 \\ \hline
com.unity3d & Game Engine & 11.8 \\ \hline
org.fmod & Game Engine & 9.6 \\ \hline
com.google.firebase & Development & 9.0 \\ \hline

 \multicolumn{3}{| c |}{\textbf{Chinese Markets} }      \\    \hline    
  com.google.ads         & Advertisement     & 25.7 \\ \hline
  org.apache        & Development         & 24.1  \\ \hline
  com.google.android.gms & Development       & 20.5  \\ \hline
  \textbf{com.tencent.mm}        & Social Networking & 17.3 \\ \hline
  \textbf{com.baidu}        & Development, Map & 16.9 \\ \hline
  \textbf{com.umeng}    & Analytics, Advertisement & 16.5  \\ \hline
 com.google.gson & Development & 16.3 \\ \hline
 
  \textbf{com.alipay}        & Payment       & 11.0  \\ \hline
  com.facebook        & Social Networking       & 10.7  \\ \hline
  com.nostra13 & Development & 10.6 \\ \hline
\end{tabular}
\vspace{-0.1in}
\label{table:toplibraries}
\end{table}

Although existing studies have created several
tools or datasets for third-party library detection~\cite{libradar,ReliableLibrary,razaghpanah2015haystack,li2016investigation}, they are either too old or incomplete to fulfill our research purpose in this paper. For example, LibRadar~\cite{libradargit,libradar} is
a widely used and obfuscation-resilient tool to identify third-party libraries used in Android apps. However, it was created in 2016, and it relies on a feature dataset of libraries extracted from Google Play apps. Considering that our apps are crawled in August 2017, and most of our apps are from Chinese markets, it may fail to report new libraries (or new versions) as well as libraries specific to the Chinese market.

To this end, we have applied the clustering-based approach introduced in LibRadar~\cite{libradar} to the 6 million apps we collected in this paper, and build a new and complete feature dataset of third-party libraries covering apps in both Google Play and the Chinese markets. At a result, we have created a dataset containing 5,102 libraries with 672,052 different versions.
We then manually examined the top 2,000 libraries and labeled them into different categories\footnote{Note that for libraries with multiple versions, we only need to label one of them.}. In order to identify
the company behind each one of them as well as the purpose of the library, 
we search the unobfuscated package name in Google, and refer to several sources, including AppBrain library classification~\cite{appbrain}, PrivacyGrade classification~\cite{privacygrade}, and Common Library classification~\cite{li2016investigation}. We group them
in 5 different categories by their purpose or offered service:
ad network, analytics, social networking, development tools, and payment. 

As shown in Figure~\ref{fig:tpl}, 
the presence of third-party services varies
from app store to app store, yet it remains high:
Google Play has the highest presence of embedded third-party libraries
in their published apps (roughly 94\% of published apps have
a third-party library) whereas PC Online
presents the lowest penetration (85\% of published apps). 
Differences also appear in terms of the total number of 
libraries per app when inspecting specific stores, 
specially in vendor-provided
app stores. While the average app embeds more than 10 
third-party libraries, those published in 360 market
have 20 third-party libraries embedded on average. 
This number contrasts significantly 
with the 8 libraries found in average for Google Play apps. 

\noindent {\textbf{Most Popular Third-party Libraries.}} 
Table~\ref{table:toplibraries} lists the top 10 third-party 
libraries found for apps published in Google Play and all 
Chinese markets, respectively. 
Google-related libraries used for advertisement and analytics services dominate in Google Play: they can be found in more than 60\% of 
published apps. 
It is interesting to see that, although Google services are blocked in China, Google-related libraries can be also found in Chinese markets, with more than a quarter of apps in Chinese markets embedding Google-related advertising libraries~\cite{vallina2012breaking,razaghpanah2018apps}. We further explored these apps and identified two leading reasons for this. The first reason is that most of these apps do not release Chinese-specific versions. This
implies that the subset of applications relying on Google Services found
on alternative Chinese app stores are identical to those present in Google Play. The second reason is that some markets crawl apps in Google Play to enlarge their application catalog: more than 30,000 apps 
published in Baidu market are explicitly labeled as crawled 
from Google Play in the developer name field. 
Nevertheless, we found many instances of third-party libraries
specific to the Chinese market across app stores. 
For instance, instead of Facebook's GRAPH API~\cite{facebookapi}, 
more than 17.3\% of the apps published in Chinese markets embed 
Tencent Wechat library~\cite{wechatSDK},  
a popular Chinese social networking SDK. 
Alipay (a payment SDK) and Baidu (a library for development also 
offering map support) are also used by more than 10\% of the apps
published in Chinese markets, hence replacing Google 
vending and Google Maps, respectively.

\noindent {\textbf{Advertising libraries.}} 
Identifying advertising libraries is a non-trivial task, as suggested by previous studies~\cite{PEDAL, ADDetect, Dong-FSE-18}. We leverage AppBrain and Common Library classification~\cite{li2016investigation} to identify and
classify third-party ad libraries. 
We have manually labeled
282 advertisement-specific libraries (with 56,011 versions) in total. 
As shown in Figure~\ref{fig:ad}, around 70\% of the apps published in Google Play use any of the labeled ad libraries, while 53.2\% of apps in Chinese markets use at least one ad library. 
It is worth mentioning that Google AdMob dominates Google Play with roughly 90\% of the advertisement market share, while the Chinese mobile ad ecosystem is more decentralized. Google AdMob and Umeng are the two most popular ad libraries, accounting for 80\% of the mobile ad market share in China, while more than 200 ad libraries compete for the remaining 20\% of the market.

\begin{figure}[t]
  \centering
  \includegraphics[width=3.4in]{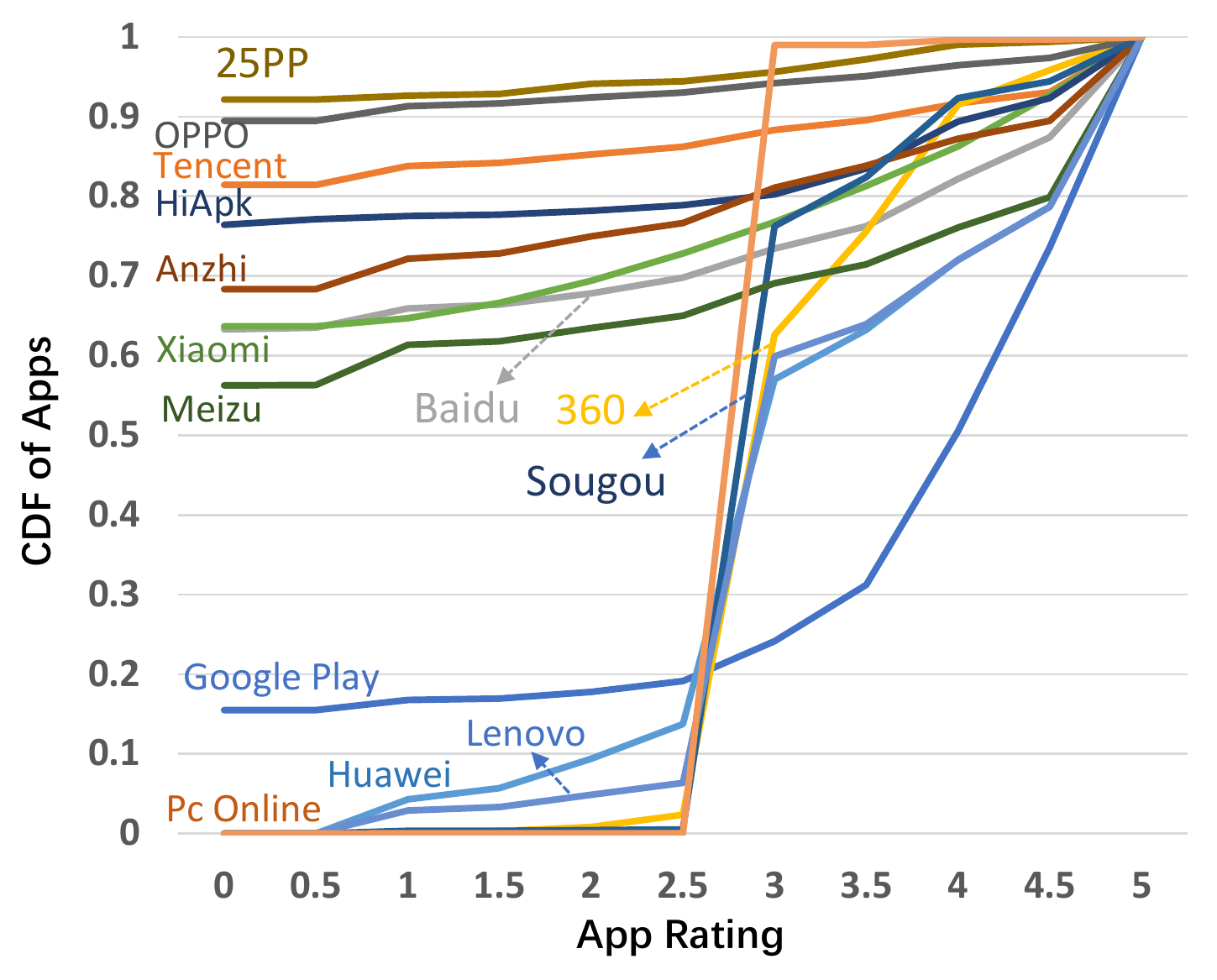}
  \vspace{-0.15in}
  \caption{Distribution of app ratings across markets. 
  }
    \vspace{-0.15in}
  \label{fig:ratingDistribution}
\end{figure}

\subsection{App Ratings}

We conclude our app store comparison 
with a brief analysis of how users rate published apps. The rating scores are crawled from app markets. Note that if an app does not
receive any rating score, we set it as 0 by default. Figure~\ref{fig:ratingDistribution} shows the CDF of app ratings for all the considered markets. 
The distribution shows that app ratings vary greatly across Chinese markets but it is possible to identify two clear patterns:

\begin{itemize}[leftmargin=*]
\item \textbf{Pattern \#1:} More than 80\% of apps in Chinese third-party app markets do not receive user ratings at all, around 90\% of these apps have less
than 1,000 downloads. This pattern can be found in
25PP, OPPO and Tencent Myapp markets. This trend
indicates that most of the apps published in these markets are low-quality and unpopular Android apps -- a trend in-line with the app download distribution shown in Figure~\ref{fig:downloadsDistribution}.
\item \textbf{Pattern \#2:} Finally, we notice that the distribution for 
several markets (\eg PC Online in the bottom) contains many apps 
with ratings between 2.5 and 3 out of 5. We tried to upload some testing 
apps to PC Online and found that they use a default rating of 3, 
instead of a default rating 0, which could be the reason leading 
to this distribution. 
\end{itemize}

\noindent Google Play, instead, presents a pattern
completely different to that of any Chinese app market: 
only 9.3\% of Google Play apps have not been rated 
by users, while more than 50\% of them have received ratings 
higher than 4.

\section{Publishing dynamics}
\label{sec:publishing}

In this section, we investigate the publishing dynamics of app developers.
We focus on analyzing
the publishing distribution for each developer and app across each store\footnote{We identify unique apps across markets based on their package names
(or app ID).}.
We define ``\textbf{single-store}'' released apps as those available only in a single market of our dataset;
otherwise, we label them as ``\textbf{multi-store}'' apps. Note that it is possible that the ``single-store'' app would appear in other markets that are not covered in this paper. This, however, does not affect our 
comparative study.

\begin{figure}[t]
\small
  \centering
  \includegraphics[width=\linewidth]{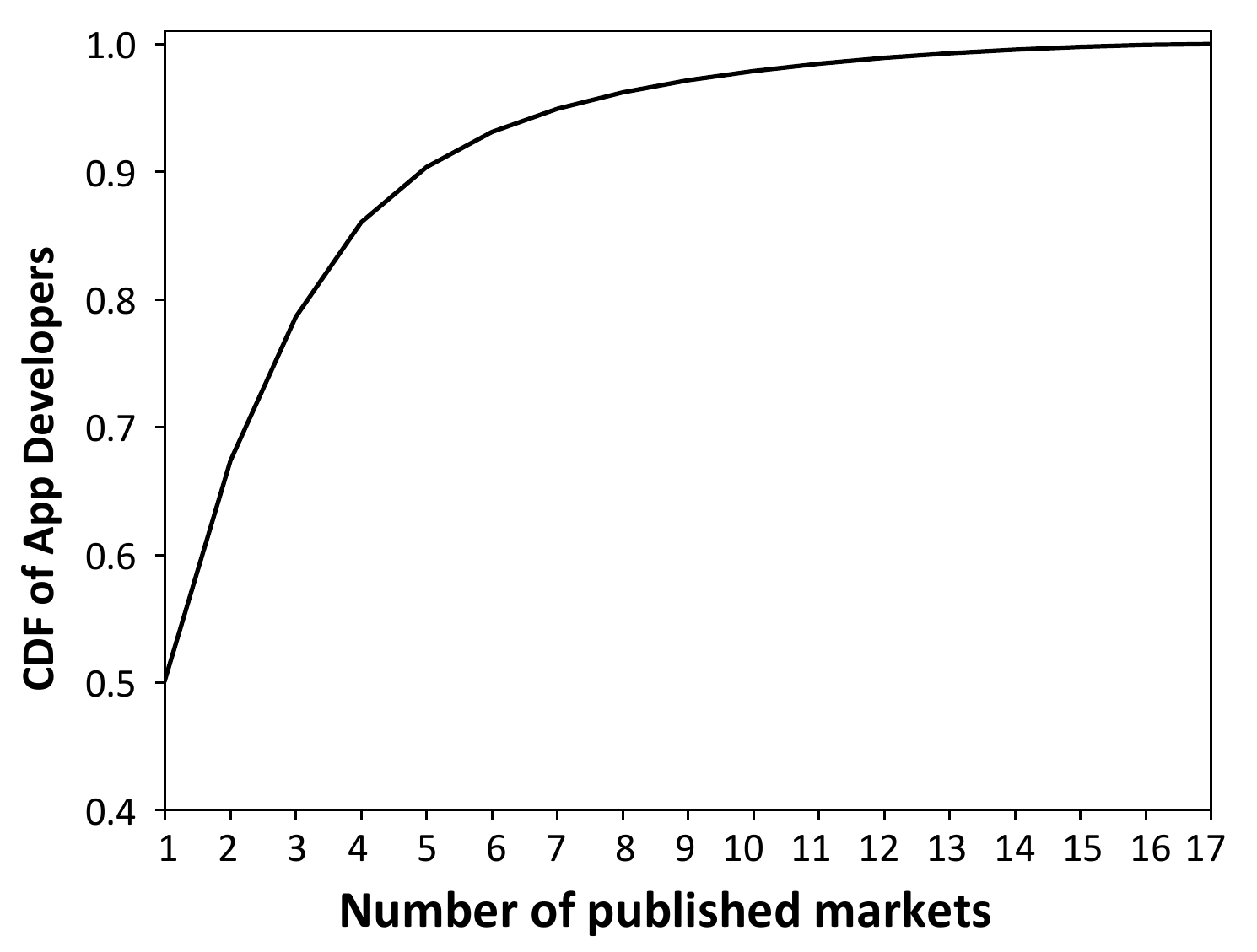}
    \vspace{-0.1in}
  \caption{CDF of developer published markets.}
  \vspace{-0.15in}
  \label{fig:marketnumber}
\end{figure}

\begin{figure*}
\centering
\subfigure[CDF of Apps VS. Number of App Versions]{\label{fig:cdf1}
\includegraphics[width=0.65\columnwidth]{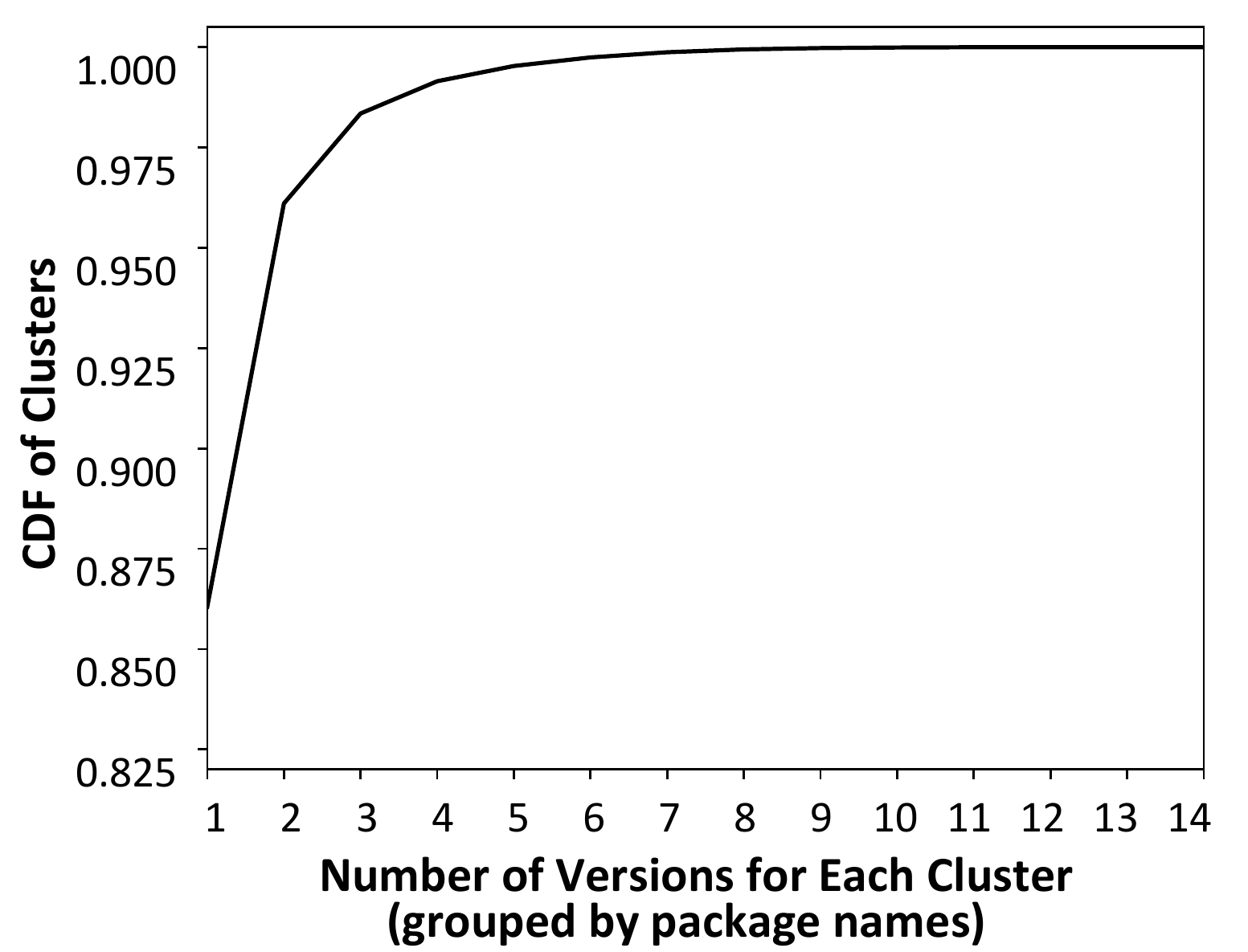}
}
\subfigure[CDF of Apps VS. Cluster Size]{\label{fig:cdf2}
\includegraphics[width=0.65\columnwidth]{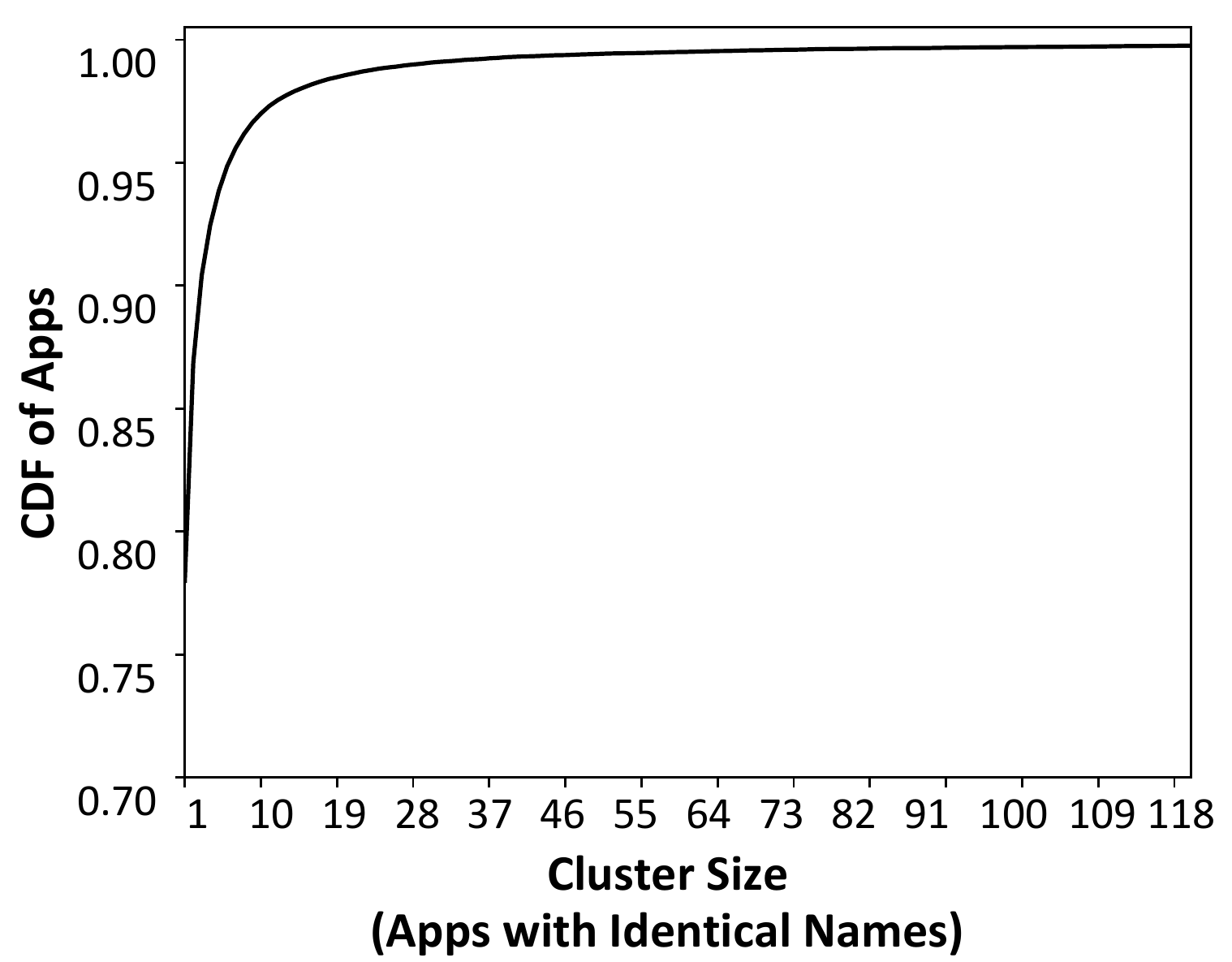}
}
\subfigure[CDF of Apps VS. Number of Developers]{\label{fig:cdf3}
\includegraphics[width=0.65\columnwidth]{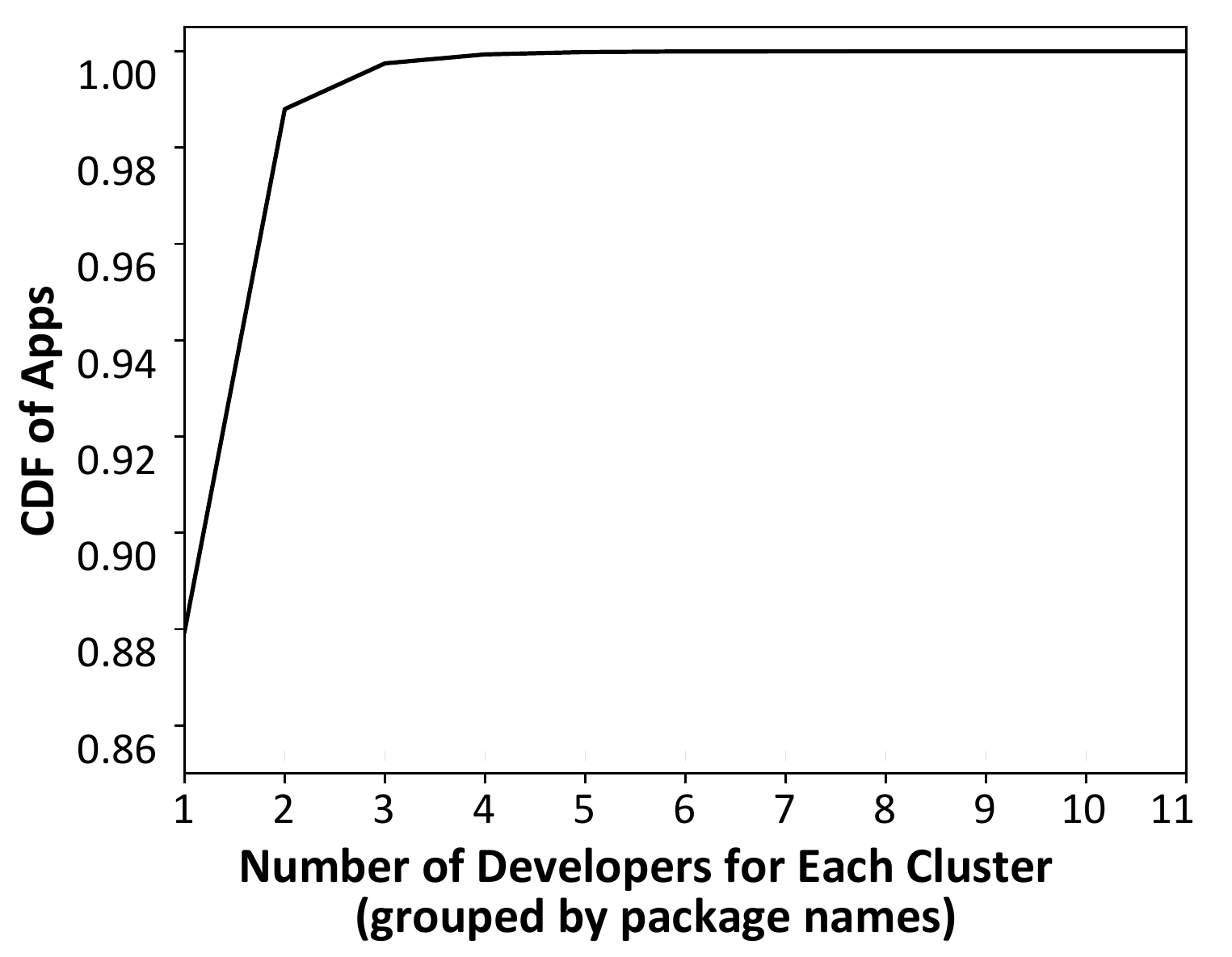}
}
\vspace{-0.1in}
\caption{CDF of apps vs. (a) number of different versions (b) cluster size, and (c) number of developers.}
\vspace{-0.1in}
\label{fig:CDFAll}
\end{figure*}

\subsection{App Developers}
\label{sec:developers}

Android mobile apps must be signed with a developer key before being released.
We used the tool ApkSigner~\cite{apkSigner}
to extract the app developer signature present on each APK.
This metadata, embedded on each executable, cannot be spoofed or 
modified by malicious actors\footnote{We found that one
developer (with the same signature) may correspond to multiple names across markets with slight variations, \eg Chinese name vs. 
English name.}.
We identified slightly over 1 million app developers -- all of them with different signatures -- in our dataset, as summarized in Table~\ref{table:marketfeature}.

Our analysis reveals that app developers
follow different publishing
strategies by targeting app stores and users in different ways. 
More than half of the developers release their apps in Google Play, 
and around 48\% of them
focus solely on Chinese alternative markets. 
Out of these developers found on Google Play, 57\% of them do not release their apps in Chinese markets, possibly due to language barriers or a lack of understanding of the fragmented Chinese ecosystem.

Figure~\ref{fig:marketnumber} shows the CDF of the number of app markets targeted by each app developer. Around
20\% of the app developers publish their apps in more than 3 app stores simultaneously, but 
only a few of them (just 696) 
roll out their apps in the 17 markets simultaneously. 
It is interesting to note that
over 10\% of the developers target exclusively one single  
Chinese store. This trend is more prevalent 
for those markets with a larger app catalog (\eg Tencent and 25PP), 
which also offer incentives to app developers for the 
exclusive publishing rights of their software.

\subsection{Single- and Multi-store Apps}

\noindent \textbf{Single-store Apps.}
More than 77\% of the apps published in Google Play are single-store ones. This result is expected, as Google Play has a global
presence and its catalog has far more apps than any 
other market individually. 
On average, 11\% of the apps published 
in alternative Chinese app stores are single-store, though
this figure varies across stores. For example,
while AnZhi, OPPO and 25PP have over 20\%
of single-store apps, 
both Wandoujia and Meizu markets have less than
1\% of single-released apps.
A manual inspection of the apps exclusively
published in Meizu reveals that they are popular apps 
explicitly developed for Meizu-branded handsets
(\eg {\tt com.meizu.flyme.wallet} and {\tt com.meizu.media.reader}).

\noindent \textbf{Multi-store Apps.} Between 20\% and 30\% of the apps 
published in Chinese alternative markets are also present in Google Play. The analysis also indicates that many Chinese markets share 
a significant fraction of their app catalogs: for instance,
80\% of the apps published in 25PP are also
released in Huawei, Wandoujia, Meizu and Lenovo markets. 
This trend is also present 
among the top 1\% most popular apps (by downloads) for each market:
over 80\% of the top 1\% most popular apps are shared across all
Chinese markets. Catalog similarities between top 
apps in Chinese stores and Google Play are, instead, low. 
This finding confirms that many developers target exclusively Chinese app stores.

\subsection{IDE and App Store Introduced Biases}

The previous method offers an upper-bound estimation of catalog overlaps
between stores. However, an important remaining question is: 
\textit{are two apps with the same package name and app version identical?} 
An alternative and stricter method to identify whether two
apps are identical is comparing the hash (e.g., MD5) of their APK contents. 
This method allowed us to find a total number of 
546,703 apps in our dataset
with identical package names, version code and developer but
different MD5.
For instance, we have 14 different hashes for
the app {\tt com.kugou.android} v8.7.0.
After manually inspecting their DEX files (\ie main function code),
we conclude that those apps are identical:
the only difference between them is 
their \texttt{META-INF/kgchannel} file\footnote{The META-INF/kgchannel files/directories are
created, recognized and interpreted by the Java 2 Platform to configure apps, 
extensions, class loaders and services. 
The main purpose of them is to differentiate the source of app users 
(\ie from which market the app is installed).}.
This confirms that relying on the package name, 
version number and developer signature are sufficient to 
accurately identify similar apps despite these subtle differences. 
Finally, we also identify instances of app store-introduced 
differences resulting from stores forcing app developers
to follow certain requirements prior to publication. A notable case is 
360 market, which requires developers to 
obfuscate their apps with 360 Jiagubao before uploading 
it to the app store~\cite{360jiagu}.

\subsection{Outdated Apps}

\begin{figure}[th]
\small
  \centering
  \includegraphics[width=3.4in]{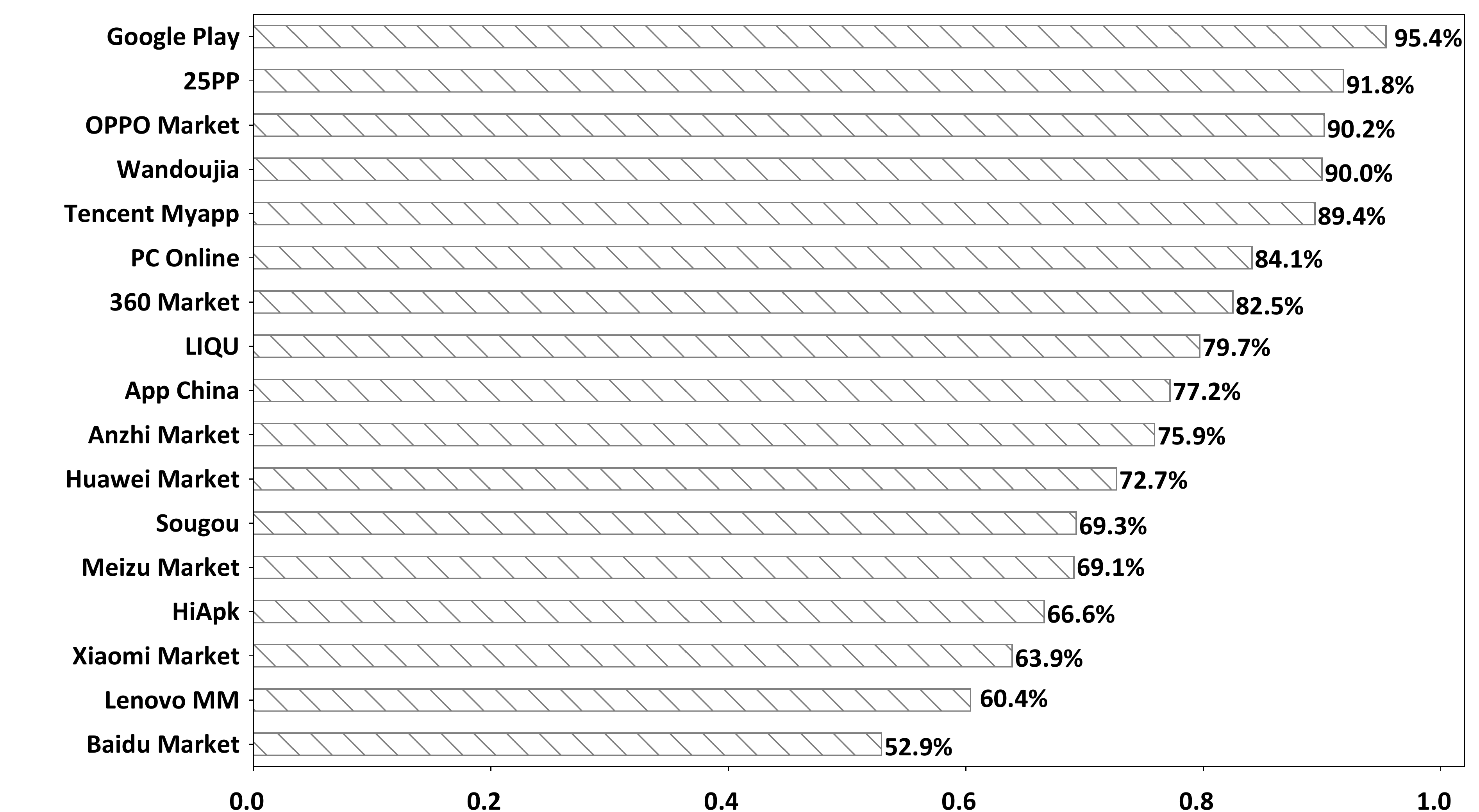}
\vspace{-0.15in}
  \caption{A comparison of app updates across markets.}
  \vspace{-0.1in}
  \label{fig:version}
\end{figure}

Another reason potentially preventing us from  
identifying multi-store released apps are unsynchronized 
roll-outs of new app versions across stores. 
We now relax the condition
to identify two identical apps: we only consider
the app package name and developer signature, excluding the 
app version\footnote{We assume that app version numbers are assigned incrementally regardless of app stores.}. As shown in Figure~\ref{fig:CDFAll}(a), 
roughly 14\% of apps have simultaneously published 
multiple versions in different stores, up to 14 different versions in
extreme cases.
Because we use a ``parallel search'' strategy in our crawler (Section~\ref{sec:crawling}), 
the elapsed time between all crawls for a given app across markets is in the
order of a few minutes, so those are intentional actions or poor 
software maintenance practices of the developers.
This behavior is not limited to poorly
maintained unpopular apps.

Figure~\ref{fig:version} details the overall distribution of outdated apps
across app stores. Note that for this analysis
we exclude single-store apps which are always updated by definition. 
Besides unfixed bugs and potential vulnerabilities, publicly available 
outdated apps also hinder users from enjoying newly added features. 
This can decrease the 
perceived quality of the apps, and overall hurts the brand equity of the market. 
This observation suggests that developers may prioritize roll-outs 
in specific app stores. 
Google Play has
the highest version number across all app stores:  
95.4\% of the apps published there have the highest app version number. 
This is not the case for stores like Lenovo MM and Baidu markets,
where more than 39\% of their apps might be outdated
according to their version number.

\section{Developer Misbehaviors}
\label{sec:misbehaviors}

In this section, we study the prevalence of various types of malicious behaviors 
across markets. Specifically, we study the presence of
\emph{fake apps}, \emph{cloned apps}, \emph{over-privileged apps}, and
\emph{malware}. 
The differences between fake and cloned apps are subtle but
substantial. 
Malicious developers can release fake apps that masquerade
as the legitimate one but stealthily perform malicious
actions on the user's device. We define those as ``fake apps''
~\cite{zhou2012dissecting}. We consider ``cloned apps'' as those
that are the results of repackaging legitimate ones~\cite{wukong}. 

\begin{table}[t]
\small
\newcommand{\tabincell}[2]{\begin{tabular}{@{}#1@{}}#2\end{tabular}}
\centering
\caption{Fake and cloned apps across stores. SB and CB stand for Signature-Based and
Code-Based clones, respectively.}
    \vspace{-0.05in}
\label{table:fakeclone}
\begin{tabular}{lrrr}
\hline
& & \multicolumn{2}{c}{Clones} \\ \cline{3-4}
\textbf{Market} & \textbf{Fake (\%) } & \textbf{SB (\%)} & \textbf{CB (\%)} \\
\hline
Google Play & 0.03 & 4.01 & 17.82 \\
Tencent Myapp & 0.53 & 8.24 & 22.73 \\
Baidu Market & 0.48 & \textbf{10.98} & 17.38 \\
360 Market & 0.50 & 5.43 & \textbf{23.26} \\
Huawei Market & 0.33 & \textbf{11.54} & 18.76 \\
Xiaomi Market & 0.0 & 8.00 & 20.11 \\
Wandoujia & 0.39 & 5.98 & 21.23 \\
HiApk &  0.64 &  7.51 & 20.08 \\
AnZhi Market & 0.57 &  4.92 & 20.71 \\
OPPO Market &  0.38 & 5.85 & 20.94 \\
25PP &  0.35 & 7.16 & \textbf{24.08} \\
Sougou & \textbf{1.83} &  4.86 & 18.28 \\
MeiZu Market & \textbf{1.14} & 6.65 & 18.42 \\
LIQU &  0.40 & 5.32 & 16.68 \\
App China & 0.0 & \textbf{10.17} & 13.23 \\
Lenovo MM & 0.67 & 7.81 & 16.37 \\
PC Online & \textbf{1.89} & 8.60 & \textbf{23.34} \\ \hline
Average & 0.60 & 7.24 & 19.61 \\
\hline
\end{tabular}
    \vspace{-0.15in}
\end{table}

\subsection{Fake Apps}

We exploit the fact that fake apps usually try to emulate the app name 
of a legitimate one, but are published 
with different package names~\cite{grayware, Sumon}.
We applied a clustering-based method to efficiently identify fake apps at scale.
First, we build a cluster enforcing a strict matching of app names.
As shown in Figure~\ref{fig:CDFAll}(b),
around 22\% of the apps in our dataset share the same name with at least
another app, all of them with different package names, either in the
same or in a different store. 
Not all the identified apps are necessarily fake, as developers may have
legitimate reasons for releasing different apps (package names) with the same app name.
This is the case of: 1) apps sharing common names
like \textit{Flashlight},
\textit{Calculator},  or \textit{Wallpaper}; and 2) 
apps released by the same developer with different package names
for different platforms\footnote{\eg   
\texttt{com.sogou.map.android.maps} and 
\texttt{com.sogou.map.android.maps.pad} are two different versions of \texttt{Sogou Map}.}.

To this end, we applied a heuristic rule to remove legitimate clusters. 
Generally, the apps in a cluster include different developer signatures.
By manually analyzing 100 randomly selected clusters of 
different size, we found out that 83\% of fake apps form small clusters (size $<5$ with uncommon names)
of unpopular ones (\ie
downloads $\leq 1,000$) and a popular one with more than 1 million installs (the official app). 
Table~\ref{table:fakeclone} summarizes the percentage of fake apps identified in
each market using this heuristic. 
The result suggests that fake apps are present in all app stores, including Google Play. Nevertheless, Meizu, PC Online and Sougou stores have a percentage of fake apps above the average. Note that our heuristic is straightforward yet very effective in identifying apps that use similar names to camouflage as the official apps.

The largest number of fake apps
in absolute terms correspond to 25PP and Tencent Myapp, with 3,591 and
3,347 apps, respectively. Relative to the market size, PC Online (with 1.89\%) and
SouGou (1.83\%) lead the ranking of markets with higher
presence of fake apps. 
Overall, our results
suggest that many app markets do not take enough efforts to identify and remove fake apps, 
despite all of them--but PC Online and HiApk--
requesting copyright checks and performing app auditing before publication (Section~\ref{sec:features}).
While we did not identify any fake app in Xiaomi and App China,
Google Play presents a marginal number of fake apps (572 in total). 

\subsection{Cloned Apps}
\label{sec:clone}

Cloned apps often share a large portion of the metadata with the original app,
but they are obviously signed by different developers. 
We explored the prevalence of
cloned apps using two separate strategies: a signature-based approach  (which aims
at identifying apps with the same package name but different developer signatures)
and a code-based approach (\ie apps with high code similarity but different
package names).
However, we are also interested in identifying 
the source market in which the original app has been published.
As it is non-trivial to identify the original app given a pair of
cloned apps~\cite{Piggyback, AdRob}, we
resort to a heuristic approach to solve this:
the app with more downloads is regarded as the original one.
Unfortunately, this may generate
false positives as 
it may be possible for the cloned app
to have more installs than its original version. 
Unfortunately, to the best of our knowledge
the research community has not developed a more accurate method to solve this
problem~\cite{wukong, Piggyback, DNADroid}.

\noindent\textbf{Signature-based clones.}
As in the previous section, we first cluster
all the apps by their package name and then compare the app developer signatures
for each cluster. 
We consider that two apps are clones if they share the same
package name but do not have a common developer signature. Since package names are
supposed to distinctively identify an Android app, it is expected that they
should be unique across different Android markets and that they are signed with the same developer key. 

Figure~\ref{fig:CDFAll}(c) shows the distribution of apps with respect to the number
of developer signatures obtained in a cluster. Overall, 12\% of apps have at least 2 
clones released by different developers. 
For example, the app \texttt{com.dino.dinosuperapp} has been published in 
15 different markets by 11 different developers. To better understand
the nature of these clones, we manually examined 100 randomly selected pairs of
signature-based clones. In all cases, we observed that clones are actually repackaged
apps, \ie apps created by disassembling the original app, making modifications, and
finally reassembling the resulting code into a different app. Even if we cannot cover all cases manually, our analysis suggests that there are no legitimate reasons behind these identified clones.

\noindent\textbf{Code-based clones.}
Since cloned apps can also modify the package name, we implemented a different
approach based on analyzing code similarity to identify cloned apps. Previous work has
proposed different approaches for app repackaging 
detection~\cite{DroidMoss,viewdroid,Juxtapp,AdDarwin,DNADroid,Piggyback}. 
In this paper, our implementation is based on WuKong~\cite{wukong}, which 
proved to be an accurate and scalable two-phase approach for app clone detection.
We first extracted Android API calls, Intents, and Content Providers for each app and created a feature vector per app with more than 45K dimensions. We then used a variant of the Manhattan
distance to measure the similarity between each pair of vectors. Specifically, for
$n$-dimensional feature vectors $A$ and $B$, their distance is given by
\[distance(A, B) = \frac{\sum_{i=0}^{n}\left | A_{i} - B_{i}\right |}{\sum_{i=0}^{n}\left ( A_{i} + B_{i} \right )}.\]
If the resulting distance between the computed vectors for a pair of apps exceeds a certain threshold -- we experimentally
selected a conservative threshold of 0.05, which corresponds
to a 95\% similarity -- 
and they are signed with different signatures, we consider 
these two apps as potential clones. 
For those apps flagged as potential app clones, we performed a second 
code-level comparison to refine the results as introduced by WuKong. 
In this second step, we consider two apps to be clones 
when they share more than 85\% of the 
code segments. Due to space limitations, we omit the implementation details here.

\begin{figure}[t]
\small
  \centering
  \includegraphics[width=3.25in]{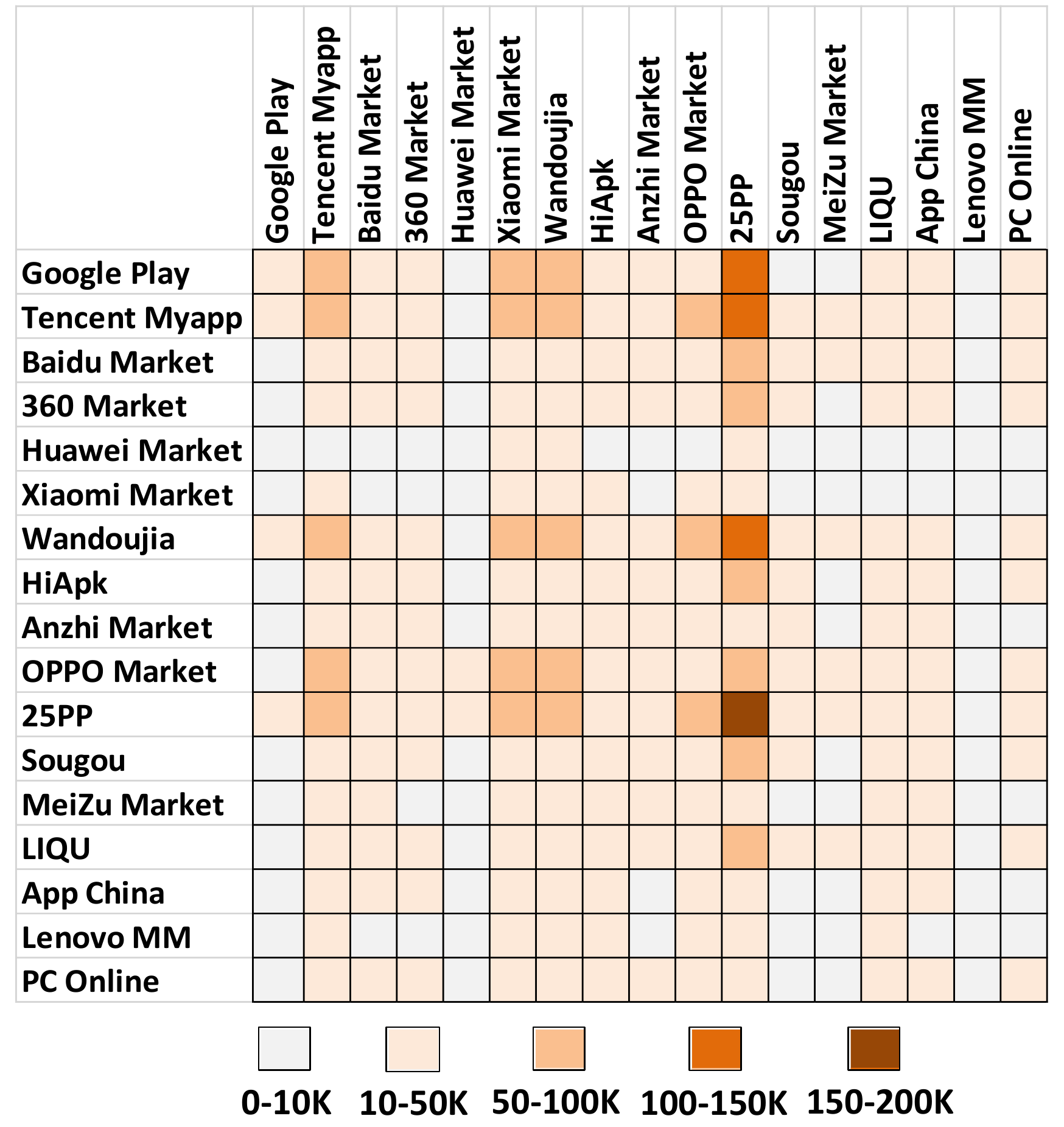}
    \vspace{-0.1in}
  \caption{Intra- and inter-market app clones.}
  \vspace{-0.2in}
  \label{fig:repackage}
\end{figure}

Previous work has suggested that, on average, more than 60\% of an app's code come from third-party libraries~\cite{wukong}. This figure is relevant for our analysis 
since libraries may cause both false positives and false negatives when detecting code clones~\cite{li2016investigation}.
To overcome this limitation, we leveraged LibRadar~\cite{libradar,libradargit} 
to identify and eliminate the impact of third-party libraries on our code-based app clone study.

\noindent\textbf{Results.} Table~\ref{table:fakeclone} summarizes the distribution of signature- and
code-based clones for each market. Code-based clones (roughly 20\%) are generally more
common than signature-based clones (roughly 10\%).
This result is in line with figures reported in previous work~\cite{wukong,DroidMoss} and
suggests that attackers are more interested in advanced cloning 
methods that go beyond changing app package names and manipulating the code.
We further illustrate the source market of cloned cases in the heatmap 
rendered in 
Figure~\ref{fig:repackage}. Both intra-market and inter-market clones 
are considered~\footnote{Only code-based clones are presented as signature-based 
clones do not involve any intra-market clones.}.
For each cell (row X, column Y), the color represents the number of cloned apps in market Y that were originally published in market X.
Google Play is the premier source for cloning apps: it presents the
large number of apps being cloned into Chinese markets. We can also detect
interesting trends when looking at the destination of these 
apps. 
Market 25PP has the largest number of cloned apps, which are mainly copied from Google Play, 
followed by Tencent Myapp and Wandoujia.
Surprisingly, intra-market clones are also quite common in addition to inter-market clones.
As shown in Figure~\ref{fig:repackage}, more than 181,677 apps in 
25PP market have similar apps to those originally from the same market.

\subsection{Over-privileged Apps}

\begin{figure}[t!]
\small
  \centering
  \includegraphics[width=3.4in]{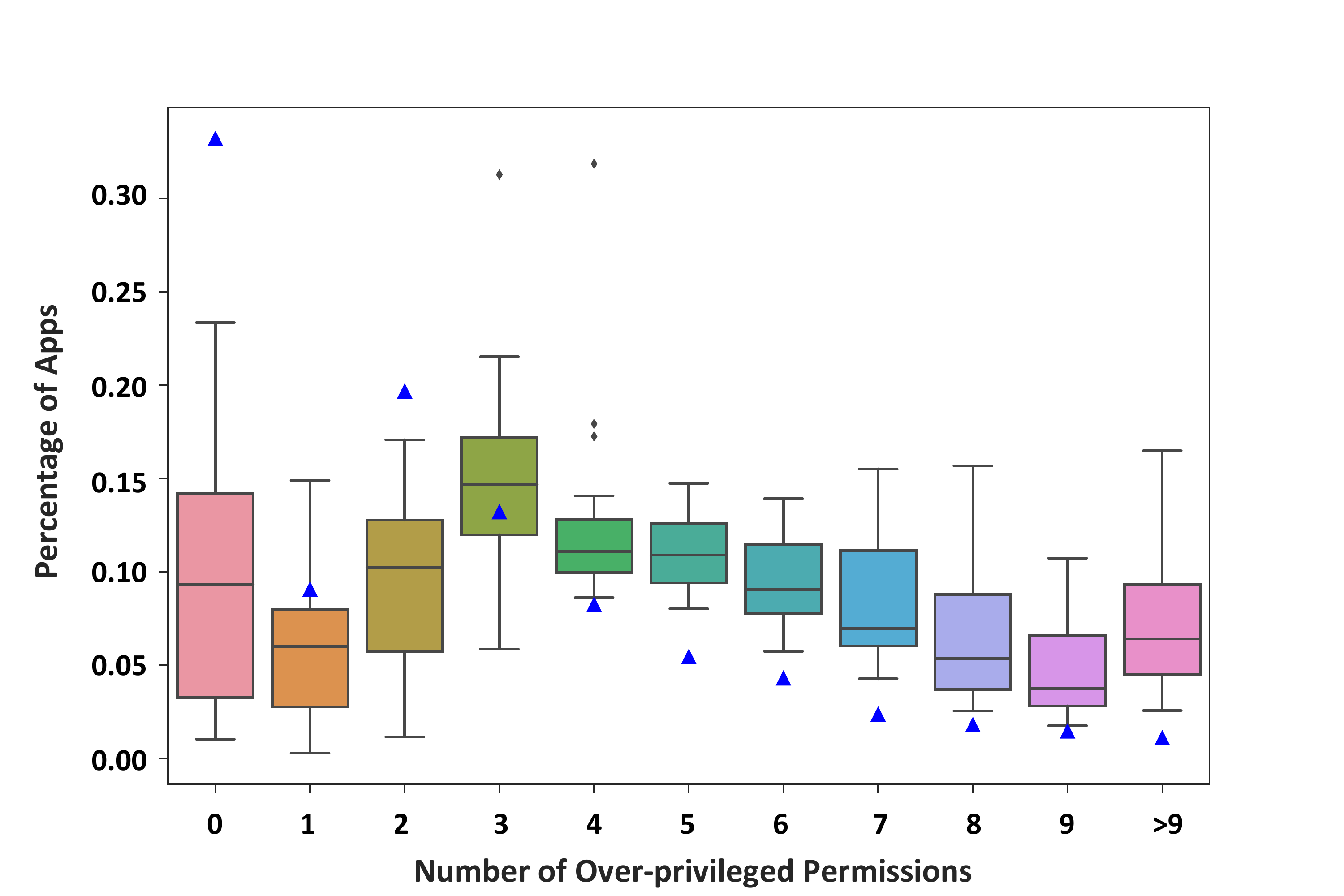}
  \caption{Distribution of over-privileged apps in Google Play and Chinese markets. The triangle symbol represents the value for Google Play, while the box-plots represent the values across the 16 Chinese alternative stores. }
  \label{fig:overprivileged}
\end{figure}

Previous studies~\cite{pscout-paper, stowaway} have analyzed the gap between requested
permissions and those actually used by Android apps. An app is said to be ``\textbf{over-privileged}''
when it requests more permissions (listed in the AndroidManifest.xml) than those actually used in their functionalities. Previous
work~\cite{wu2013impact} has suggested that more than 85\% of Android apps published in vendor-customized
phones suffer from this issue. Since permissions constitute an explicit declaration of 
what sensitive resources an app will use~\cite{wang2015using, wang2017understandingPurpose}, over-privileging an app is undesirable because:
(i) it is a violation of the principle of least privilege~\cite{leastPrivilege}; (ii) it exposes users to
unnecessary permission warnings; and (iii) it increases the attack surface~\cite{bartel2012automatically} and the impact of the presence of a bug or vulnerability~\cite{stowaway}.

Intuitively, this gap can be identified first by 
building a permission map that identifies
what sensitive permissions are needed for each API call/Intent/Content Provider, and using static analysis to determine what permission-related invocations an app makes. Then, we can compare the actually used permissions by the app with the requested permissions listed in AndroidManifest.xml.
To do this,
we leveraged data provided by PScout~\cite{pscout-dataset, pscout-paper}, specifically
a list of 32,445
permission-related APIs, 97 permission-related Intents, 78 Content Providers URI Strings,
and 996 Content Provider URI fields\footnote{Note that we use the API-Permission mapping for Android 5.1.1, which may not reflect the new sensitive APIs introduced in subsequent system versions. However, more than 90\% of apps in our dataset target API levels less than 5.1.
In addition, a well-known limitation of static over-privilege app analysis is its inability to handle Java reflection and dynamic code loading~\cite{wang2015reevaluating}.}.

In general terms, apps published in Chinese markets tend to
request more sensitive permissions, \ie those labeled 
as dangerous by Google~\cite{gplaydeveloperPermissionOverview},
than Google Play apps.
Figure~\ref{fig:overprivileged} shows the distribution of over-privileged apps across
markets grouped by how many permissions in excess each app has. Note that, in general, Chinese
markets contain more over-privileged apps than Google Play. Approximately 65\% of
the apps in Google Play are over-privileged, while the percentage in Chinese markets is roughly
82\%. In two particular cases (25PP and App China), more than 95\% of apps requested at least one
unused permission. Apps often request no more than 10 unused permissions, 3 being the most
common value. The most common over-privileged sensitive permissions are READ\_PHONE\_STATE (52.38\%), 
ACCESS\_COARSE\_LOCATION (36.28\%), ACCESS\_FINE\_LOCATION (33.83\%),
and CAMERA (19.98\%).

\subsection{Malware Prevalence}

In order to investigate the presence of malicious and undesirable apps in our dataset, we uploaded all the apps to VirusTotal~\cite{VirusTotal}, an online analysis service that aggregates more than 60 anti-virus engines, which is widely adopted by the research community. 
Previous studies~\cite{arp2014drebin, wei2017deep} have suggested
that some anti-virus engines may not always report reliable results. 
In order to deal with such
potential false positives, we analyzed the results grouped by how many engines 
(\textbf{AV-rank}) flag a sample as malware.
Previous work have argued that a threshold of 10 engines
is a robust choice~\cite{arp2014drebin, ikram2016analysis, zheng2012adam}. 

\begin{table}[t!]
\small
\newcommand{\tabincell}[2]{\begin{tabular}{@{}#1@{}}#2\end{tabular}}
\centering
\caption{Percentage of apps labeled as malware in each market by AV-rank.}
\vspace{-0.05in}
\label{table:malware}
\begin{tabular}{rrrr}
\hline
	& \multicolumn{3}{c}{\textbf{AV-rank (\% apps)} } \\ \cline{2-4}
 \textbf{Market}  & \tabincell{l}{\textbf{$>=1$} } & \textbf{$>=10$ } & \textbf{$>=20$} \\
\hline

Google Play   & 17.03  & 2.09 &  0.32  \\
\hline 
Tencent Myapp   & 34.15 & 11.16 & 3.45 \\
Baidu Market  & 42.77 & 12.24 & 3.30 \\
360 Market & 41.40 & 12.35 & 3.10 \\
\hline 

OPPO Market  & 42.97 & 16.43 & 6.00 \\
Xiaomi Market  & 55.11 & 9.12 & 1.82 \\
MeiZu Market  & 51.40 & 10.70 &  3.14 \\
Huawei Market  & 57.48 & 4.71 & 0.57 \\
Lenovo MM  & 54.20 & 7.53 & 1.52 \\
\hline 

25PP  & 32.36 & 8.26 & 2.06 \\
Wandoujia  & 31.99 & 7.98 & 2.19 \\
HiApk & 41.89 & 11.12 & 2.72 \\
AnZhi Market &  55.32  & 11.37 & 2.41 \\
LIQU  &  45.91  & 13.00 & 4.27 \\
PC Online   & 55.93  & 24.01 & 8.37 \\
Sougou  &  52.41  & 16.53 & 4.59  \\
App China &  48.55  & 14.13 & 4.27  \\ \hline
Average &  36.49  & 12.30  & 3.69   \\
\hline 
\end{tabular}
\vspace{-0.2in}
\end{table}

\textbf{Overall Result.}
Table~\ref{table:malware} shows the overall detection results. Remarkably, roughly 50\% of the apps in Chinese markets are flagged at least by one anti-virus engine, while the percentage for Google Play
is considerably lower (17.03\%). 
According to the threshold of ``$AV-rank \geq 10$'', around 2\% of the apps in Google Play are labeled as malware, while
the percentage in Chinese markets is much higher. In fact, for 11 out of
the 16 Chinese markets the percentage of malware exceeds 10\%. A particularly remarkable case is the PC
Online market, with more than 24\% of its apps labeled as potentially
malicious. 
In absolute terms, Tencent and 25PP markets host the largest number of malicious apps (70,988 and 83,655, respectively).
On the opposite side, we find Huawei's market, with a figure (4.71\%) comparable in magnitude to that of Google Play (2.09\%).

\begin{table}[t]
\small
\newcommand{\tabincell}[2]{\begin{tabular}{@{}#1@{}}#2\end{tabular}}
\centering
\caption{Top 10 malicious apps by their AV-rank.}
\vspace{-0.1in}
\label{table:topmalware}
\resizebox{1\linewidth}{!}{
\begin{tabular}{rrrr}
\hline
\textbf{\tabincell{c}{Package Name\\(malware family)}} & \textbf{AV-Rank} & \textbf{Markets} \\
\hline
\tabincell{r}{com.trustport.mobilesecurity\\\_eicar\_test\_file (eicar)} & 48 & Wandoujia, 25PP \\ \hline
games.hexalab.home (mofin)  & 47 &LIQU \\ \hline
com.wb.gc.ljfk.baidu (ramnit)  & 47 & Baidu, HiAPK\\ \hline
com.ypt.merchant (ramnit)  & 46 & \tabincell{c}{Tencent, Wandoujia,\\OPPO, 25PP, LIQU} \\ \hline
com.wsljtwinmobi (ramnit) & 46 & Tencent, 25PP \\ \hline
com.wb.gc.ljfk.tx (ramnit) & 45 & Tencent  \\ \hline
com.wgljd (ramnit)  & 45 & Tencent, 360 \\ \hline
com.zoner.android.eicar (eicar) & 44 & \tabincell{c}{Google Play,\\ Wandoujia, 25PP} \\ \hline
com.zhiyun.cnhyb.activity (ramnit)  & 44 & Baidu\\ \hline
com.fai.shuiligongcheng (ramnit) & 44 & 25PP\\
\hline
\end{tabular}}
\vspace{-0.1in}
\end{table}

\textbf{Top Malware.}
Table~\ref{table:topmalware} lists the top 10 malicious apps according to their AV-Rank.
Note that two of them (\texttt{com.trustport.mobilesecurity\_eicar\_test\_file} 
and
\texttt{com.zoner.android.eicar})
correspond to the AV benchmarking
apps developed by the European Institute for Computer Antivirus Research (EICAR).
The remaining apps--and others that were manually inspected by us--clearly show potentially
malicious behaviors. For example, \texttt{com.ypt.merchant}, published in 5 markets, poses itself as a legitimate
mobile point-of-sale (mPOS) for merchants and individuals.

\textbf{Repackaged Malware.} The Android Genome Project~\cite{zhou2012dissecting} suggested that app repackaging is the main way for malware distribution, and 86\% of the 1,260 samples are repackaged malware. 
However, this dataset is outdated (collected in 2011) and the number of samples is relatively small so 
it may no longer provide a representative picture
of the current Android malware landscape.
Thus, we further analyzed how many malware samples in our dataset are repackaged apps. 
To this end, we merged the malware results with the app clone detection results as shown in Section~\ref{sec:clone}, and observed that only 38.3\% of these malware samples are repackaged apps. This result suggests that app repackaging is no longer the main way for malware spreading. We believe this is an interesting observation for our community, and we leave to future work analyzing the newest trends in malware spreading strategies.

\begin{figure}[t!]
  \centering
  \includegraphics[width=3.3in]{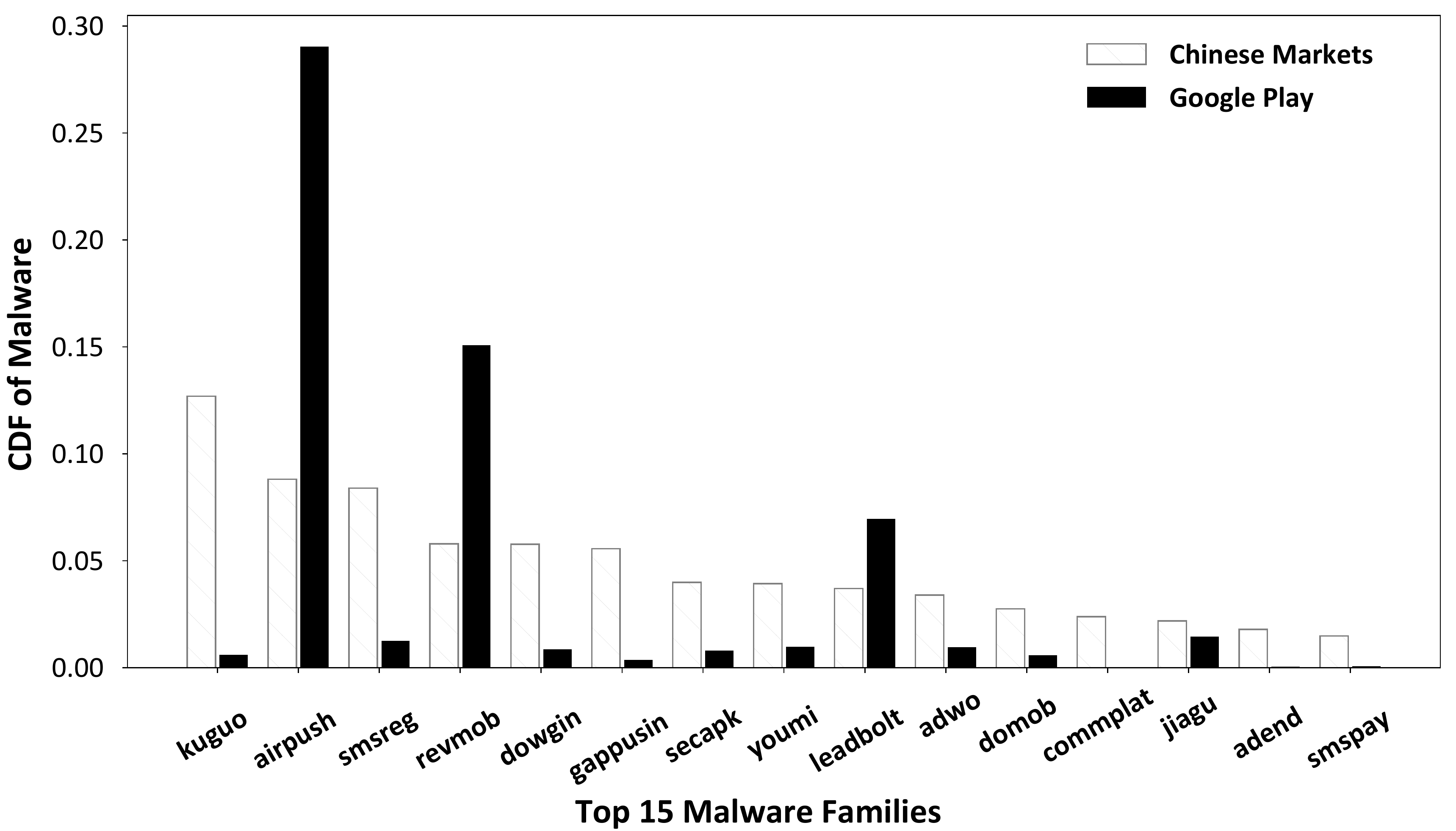}
    \vspace{-0.05in}
  \caption{Distribution of top 15 malware families in Google Play and Chinese markets.}
  \label{fig:malwarefamily}
\end{figure}

\textbf{Malware Family.}
We further analyzed the distribution of malware families across 
Google Play and Chinese markets. To do this, we used AVClass~\cite{AVclass} 
to obtain the family name (label) of each identified malware. Figure~\ref{fig:malwarefamily} 
shows the distribution top 20 malware families. An interesting finding is that
the distribution of malware families differs greatly between Google Play and Chinese markets. 
The most popular malware family in Chinese markets is \textbf{kuguo} (12.69\%), 
while it only corresponds to 0.6\% of malware in Google Play. Roughly, 45\% of the malware present in
Google Play belong to the family \textbf{airpush} (29.04\%) and \textbf{revmob} (15.09\%).
We further enlarged our threshold to ``$AV-rank \geq 20$'' and found that it shows generally similar malware family distribution.

\section{Post-analysis}
\label{sec:post}

\begin{table}[t]
\small
\newcommand{\tabincell}[2]{\begin{tabular}{@{}#1@{}}#2\end{tabular}}
\centering
\caption{Percentage of removed malware across markets. The third column indicates the number of apps also published and removed in Google Play (GPRM).}
\begin{tabular}{rrrr}
\hline
\textbf{Market} & \textbf{\tabincell{c}{\%Malware\\ Removed}} & \textbf{\tabincell{c}{\#Overlapped\\with GPRM}} & \textbf{\%Removed} \\
\hline
Google Play & 84\% & - & - \\
\hline
Tencent Myapp & 8.75\% & 7,157 & 3.1\%\\
Baidu Market & 23.99\%  & 1,422 & 34.53\% \\
360  & 43\% & 1,198 & 34.22\% \\
Xiaomi & 32.50\% & 636 & 31.13\% \\
Meizu & 29.18\% & 668 & 26.20\% \\
Huawei  & 26.92\% & 169 & 23.08\% \\
Lenovo MM & 22.75\% & 263 & 16.35\%\\
25PP &  19.63\% & 7,804 & 17.31\%\\
Wandoujia & 34.51\%  & 5,289 & 44.74\%  \\
AnZhi & 27.61\% & 632 & 25.78\% \\
LIQU & 14.08\% & 1,878 & 11.18\% \\
PC Online  & 0.01\% & 1,117 & 0.00\% \\
Sougou & 24.24\% & 1,082 & 22.00\% \\
App China & 20.51\% & 546 & 30.24\% \\
\hline
\end{tabular}
\label{table:removedMalware}
  \vspace{-0.2in}
\end{table}

All markets have strict policies to conduct copyright and security checks
(Section~\ref{sec:features}). Yet our
results reveal that they still host a significant
amount of fake and cloned apps, as well as malware samples.
As introduced in Section~\ref{sec:crawling}, we performed a second
crawl for each app store about 8 months after 
the first one in order to quantify whether 
the stores made any effort to remove those samples from their
catalogs\footnote{We exclude HiAPk from this analysis as
it has discontinued its services by the end of 2017. In addition,
OPPO can only be accessed now using their market app.}.
As shown in the first column of Table~\ref{table:removedMalware}, over
84\% of the potential malicious apps found in Google Play have been removed.
However, the percentages of malware removal in Chinese 
alternative markets vary from 0.01\% (PC Online) to 34.51\% (Wandoujia). 
We extracted and inspected in detail those apps with an AV-rank $\geq 10$
removed from Google Play (GPRM) between our crawls.
11,623 of them were also found in at least one Chinese app store, and
over 70\% of them are still hosted by at least one 
Chinese market by the end of April 2018, as shown in the 
fourth column of Table~\ref{table:removedMalware}. 
Tencent and PC Online are clearly the Chinese stores
in which those potentially malicious apps still survive.

\section{Discussion}
\label{sec:discussion}

\begin{figure}[t!]
\small
  \centering
  \includegraphics[width=2.75in]{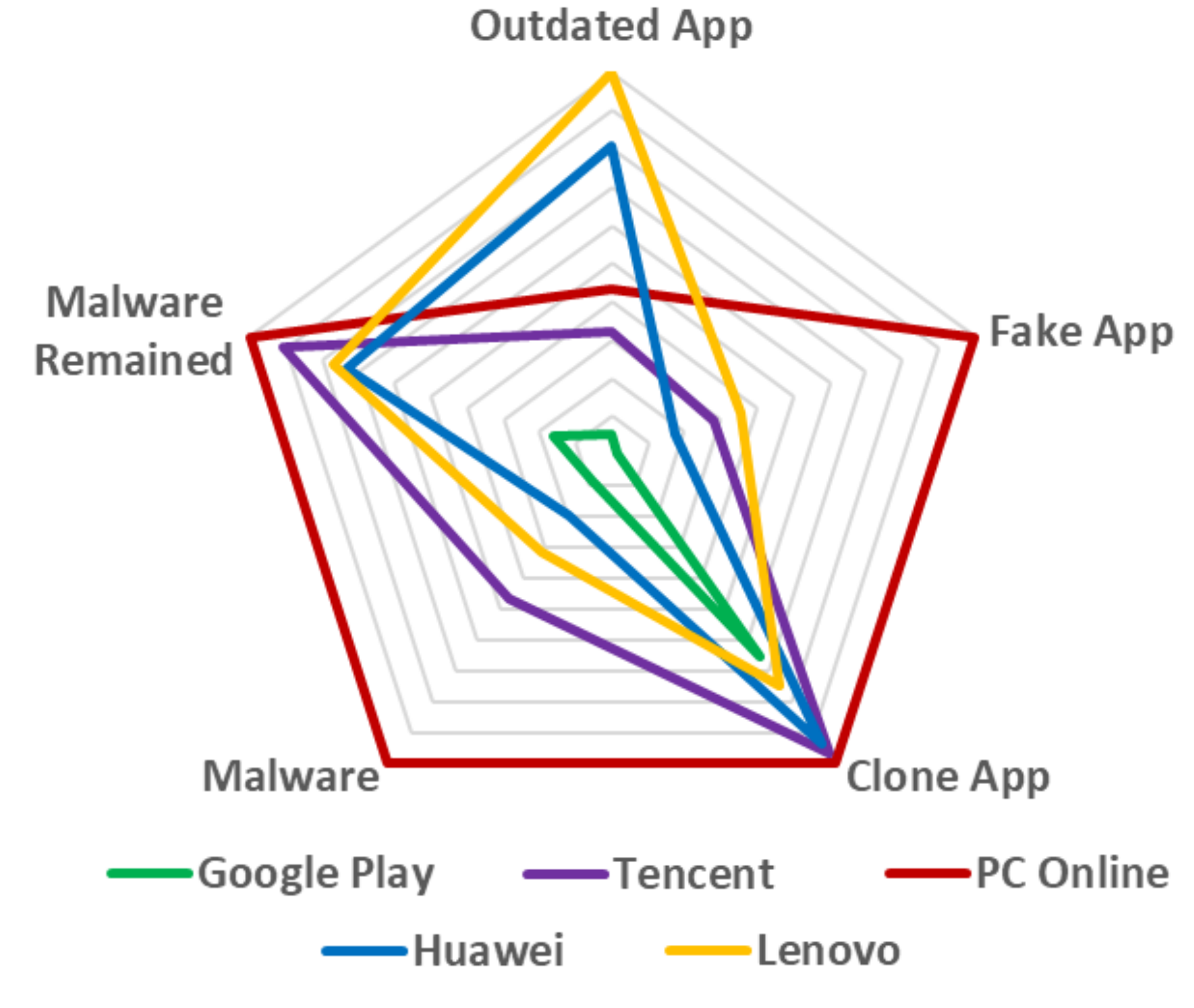}
    \vspace{-0.1in}
  \caption{Multi-dimensional comparison of Google Play, Tencent, PC Online, Huawei and Lenovo markets. For each metric, we normalize it to the scale [0, 100], and the center represents 0.}
  \vspace{-0.15in}
  \label{fig:radar}
\end{figure}

Our results reveal that potentially malicious and
deceptive activities are more common in Chinese markets than in Google Play. 
Figure~\ref{fig:radar} presents a multi-dimensional
comparison of four representative Chinese app markets and Google Play.
Tencent Myapp, one of the largest Chinese app stores by 
their aggregate number of downloads,
hosts a significant amount of mobile malware. This store seems
to be more indulgent with malicious developers, including those
publishing malware as well as fake and cloned apps. 
Although Tencent Store claims to perform manual inspection for all the submitted apps,
our empirical observations seem to contradict it. 
We find a similar behavior for PC Online. However, in this case
we could not find any developer policy describing
security checks on apps prior to publication. 

Huawei and Lenovo markets show a clearly different behavior.
These stores publish popular apps 
and present similar app ratings and download distributions.
They also seem to have strict mechanisms to prevent malware distribution:
only 4.71\% and 7.53\% of their apps, respectively were labeled as malware,
figures comparable in magnitude to that of Google Play. 
The purpose of the stores and their market segment can also
influence in the presence of malware, possibly due to their need to protect
their brand reputation. 
Lenovo's MM market does not allow individual developers to publish apps, 
a practice that could help them mitigate the spread of malware
and low-quality apps. 
However, Huawei and Lenovo markets still
have a significant number of outdated apps, which
could hinder users from enjoying newly added features and
other app improvements (\eg bug fixing). This practice
could contribute to decrease the perceived quality of the apps, hurting
as a result the brand equity of the app store.

\section{Related Work}

Previous research efforts have performed large-scale mobile app analysis~\cite{afonso2016going, chia2012app, bierma2014andlantis, bohmer2011falling, sounthiraraj2014smv, gibler2012androidleaks}.
However, alternative Android markets have not been well studied
by the research community yet.
To the best of our knowledge, our work
is the largest and most exhaustive comparative 
analysis made between the official Google Play store and 
Chinese alternative markets.

\textbf{Large-scale App Repositories.}
AndroZoo~\cite{li2017androzoo} is an academic effort focused on
compiling a large-scale dataset of APKs. This research
effort has enabled a number of studies focusing
on malicious practices and privacy risks of Android apps
~\cite{calciati2017apps, 
avdiienko2017detecting, li2015iccta, yang2017characterizing}. 
AndroZoo uses purpose-built crawlers 
to harvest more than 5M APKs from
12 app stores and 
5 Chinese markets with roughly 1.5 Million apps.
The work of Ishii \textit{et al.}~\cite{ishii2017understanding} is the closest to ours. They investigated 4.7M 
Android apps covering 27 app markets, mainly obtained from AndroZoo~\cite{androzoo},
to understand the security management of global third-party markets.

\textbf{Measurement of Google Play.}
Many research efforts have been focused on Google Play. PlayDrone~\cite{MeasurementGoogle}
also performed a large-scale characterization of 1.1 million apps 
published in Google Play. They explored various issues such as app evolution
and authentication schemes. Bogdan \etal~\cite{Longitudinal} analyzed
160,000 Google Play apps daily for a period of 6 months, aiming to summarize
the temporal patterns. Ali \etal~\cite{appleandgoogle}
quantitatively compared app market attributes (\eg ratings and prices)
of Apple store and Google Play based on 80,000 app pairs. Wang \etal~\cite{WangRemoval}
presented a large-scale study of 791,138 removed Google Play apps to
identify potential reasons for app removal. Wang \etal~\cite{Ecosystem} analyzed
the mobile app ecosystem from the perspective of app developers based on over 1.2 million apps and 320,000 developers. 

\textbf{Measurement of Alternative Markets.}
For third-party markets, Petsas \etal~\cite{MeasurementEcosystem}
analyzed 4 alternative markets to understand the downloading patterns and popularity
trends. Ng \etal~\cite{ng2014android} assessed the trustworthy level of 20 Chinese
app markets, but they only studied roughly 500 apks. Wang \etal~\cite{wang2016measuring}
have studied gaming apps across 4 Chinese markets to understand their scale and evolution. WuKong~\cite{wukong} was proposed to identify repackaged apps in five Chinese app markets.

\section{Conclusion}
In this work, we have conducted a large-scale mobile app analysis 
to understand various features of several Chinese Android app stores
and how they compare to Google Play.
Specifically, our analysis covers over
6 million Android apps obtained from 16 Chinese
app markets and Google Play. Overall, our results suggest that there are substantial
differences between the Chinese app ecosystem and Google Play, though some
minor commonalities are also found. We have identified a significant number of
developers and third-party services specialized in the Chinese market.
We have also found a higher prevalence of fake, cloned, and malicious apps in
Chinese stores than in Google Play, possibly due to market operators indulgently
oversighting copyright and security checks over the apps. 
We believe that our research efforts 
can positively contribute to bring user and developer awareness, 
attract the focus of the research community and regulators, and
promote best operational practices across app store operators.

\section*{Acknowledgments}
We sincerely thank our shepherd Prof. Zhenhua Li (Tsinghua University), and all
the anonymous reviewers for their valuable suggestions
and comments to improve this paper.
This work is supported by the National Key Research and Development Program of China (grant No.2018YFB0803603), the National Natural Science Foundation of China (grants No.61702045, and No.61772042); the BUPT Youth Research and Innovation Program (No.2017RC40); Spain's Ministry of Economy and Competitiveness (grant TIN2016-79095-C2-2-R); the Madrid Region's Technologies 2014 Research Program (grant S2013/ICE3095); the US National Science Foundation (grant CNS-1564329); and 
the European Union's Horizon 2020 Innovation Action programme (grant Agreement No. 786741, SMOOTH Project).

\bibliographystyle{abbrv}
\balance
\bibliography{sigproc}

\begin{thebibliography}{100}

\bibitem{huaweiinspection}
{Human Inspection Team in Huawei}, 2016.
\newblock \url{http://android.tgbus.com/news/bd/201609/552547.shtml}.

\bibitem{leastPrivilege}
{Principle of least privilege - Wikipedia}, 2017.
\newblock \url{https://en.wikipedia.org/wiki/Principle\_of\_least\_privilege}.

\bibitem{MarketReport5}
{The top 10 Android app stores in China in 2017}, 2017.
\newblock
  \url{https://technode.com/2017/06/02/top-10-android-app-stores-china-2017/}.

\bibitem{MarketReport1}
{2017-2018 App Market Ranking in China-iiMedia Research}, 2018.
\newblock \url{http://www.iimedia.cn/60947.html}.

\bibitem{MarketReport2}
{2018 Top 10 App Markets in China}, 2018.
\newblock \url{https://www.sohu.com/a/233552979_427024}.

\bibitem{360developer}
{360 Market - App Vetting}, 2018.
\newblock \url{http://dev.360.cn/wiki/index/id/18}.

\bibitem{360Security}
{360 Security - Free Antivirus, Booster, Cleaner}, 2018.
\newblock
  \url{https://play.google.com/store/apps/details?id=com.qihoo.security&hl=en_US
  }.

\bibitem{Alideveloper}
{Ali Platform - App Vetting}, 2018.
\newblock \url{http://aliapp.open.uc.cn/wiki/?p=140}.

\bibitem{apkSigner}
{Android Developer - APK Signer}, 2018.
\newblock \url{https://developer.android.com/studio/command-line/apksigner}.

\bibitem{gplaydeveloperPermissionOverview}
{Android Developers - Permissions Overview}, 2018.
\newblock
  \url{https://developer.android.com/guide/topics/permissions/overview}.

\bibitem{androzoo}
{AndroZoo}, 2018.
\newblock \url{https://androzoo.uni.lu/}.

\bibitem{Anzhideveloper}
{Anzhi Platform - App Vetting}, 2018.
\newblock \url{http://dev.anzhi.com/help.php?type=help&id=67}.

\bibitem{AppChinadeveloper}
{App China Platform - App Vetting}, 2018.
\newblock \url{http://dev.appchina.com/dev/help?choose=check}.

\bibitem{MarketReport4}
{App Market Ranking in China}, 2018.
\newblock \url{http://chanye.07073.com/guonei/1756627.html}.

\bibitem{baidudeveloper}
{Baidu Market - App Vetting}, 2018.
\newblock \url{http://app.baidu.com/docs?id=18&frompos=401006}.

\bibitem{gplaydeveloper}
{Developer Policy Center - Google Play}, 2018.
\newblock \url{https://play.google.com/about/developer-content-policy/}.

\bibitem{tencentdeveloper}
{Developer Policy Center - Tencent Myapp}, 2018.
\newblock
  \url{http://wiki.open.qq.com/wiki/%E5%BA%94%E7%94%A8%E4%B8%8A%E6%9E%B6%E8%A7%84%E5%88%99}.

\bibitem{facebookapi}
{Facebook Graph API}, 2018.
\newblock \url{https://developers.facebook.com/docs/graph-api}.

\bibitem{GooglePlay25Dollar}
{How to use the Play Console}, 2018.
\newblock
  \url{https://support.google.com/googleplay/android-developer/answer/6112435?hl=en}.

\bibitem{HuaweiMarket}
{Huawei has surpassed Apple as the world’s second largest smartphone brand},
  2018.
\newblock
  \url{https://www.theverge.com/2017/9/6/16259810/huawei-apple-global-smartphone-sales
  }.

\bibitem{Huaweideveloper}
{Huawei Market - App Vetting}, 2018.
\newblock \url{http://developer.huawei.com/consumer/cn/devservice/doc/50104}.

\bibitem{Lenovodeveloper}
{Lenovo Market - App Vetting}, 2018.
\newblock
  \url{http://open.lenovo.com/developer/adp/helpData/database_detail.jsp?url=http://open.lenovo.com/sdk/yysh/
  }.

\bibitem{Liqudeveloper}
{LIQU Platform - App Vetting}, 2018.
\newblock \url{http://dev.liqucn.com/User/show_article/6}.

\bibitem{Meizudeveloper}
{Meizu Market - App Vetting}, 2018.
\newblock
  \url{http://open-wiki.flyme.cn/index.php?title=%E9%AD%85%E6%97%8F%E5%BA%94%E7%94%A8%E5%95%86%E5%BA%97%E5%BA%94%E7%94%A8%E5%AE%A1%E6%A0%B8%E8%A7%84%E8%8C%83
  }.

\bibitem{OPPOdeveloper}
{OPPO Market - App Vetting}, 2018.
\newblock \url{https://open.oppomobile.com/wiki/index#id=73486}.

\bibitem{pscout-dataset}
{PScout: Analyzing the Android Permission Specification}, 2018.
\newblock \url{http://pscout.csl.toronto.edu}.

\bibitem{smartphonemarket}
{Smartphone Market in China}, 2018.
\newblock \url{https://www.statista.com/topics/1416/smartphone-market-in-china/
  }.

\bibitem{Sogoudeveloper}
{SOGOU Platform - App Vetting}, 2018.
\newblock \url{http://zhushou.sogou.com/open/question-14.html}.

\bibitem{MarketReport6}
{Top 10 Android App Stores in China}, 2018.
\newblock
  \url{https://newzoo.com/insights/rankings/top-10-android-app-stores-china/}.

\bibitem{MarketReport3}
{Top 10 Chinese App Markets}, 2018.
\newblock \url{http://www.sohu.com/a/223230295_100075221}.

\bibitem{VirusTotal}
{VirusTotal}, 2018.
\newblock \url{https://www.virustotal.com/ }.

\bibitem{wechatSDK}
{WeChat SDK}, 2018.
\newblock \url{https://open.weixin.qq.com/ }.

\bibitem{Xiaomideveloper}
{Xiaomo Market - App Vetting}, 2018.
\newblock \url{https://dev.mi.com/console/doc/detail?pId=879}.

\bibitem{360jiagu}
{360 Jiagu}, 2017.
\newblock \url{http://jiagu.360.cn}.

\bibitem{afonso2016going}
V.~M. Afonso, P.~L. de~Geus, A.~Bianchi, Y.~Fratantonio, C.~Kruegel, G.~Vigna,
  A.~Doup{\'e}, and M.~Polino.
\newblock Going native: Using a large-scale analysis of android apps to create
  a practical native-code sandboxing policy.
\newblock In {\em Proceedings of the Network and Distributed System Security
  Symposium (NDSS)}, 2016.

\bibitem{appleandgoogle}
M.~Ali, M.~E. Joorabchi, and A.~Mesbah.
\newblock Same app, different app stores: A comparative study.
\newblock In {\em Proceedings of the International Conference on Mobile
  Software Engineering and Systems (MOBILESoft)}, 2017.

\bibitem{aliyun}
{Aliyun ECS}, 2017.
\newblock \url{https://ecs-buy.aliyun.com}.

\bibitem{grayware}
B.~Andow, A.~Nadkarni, B.~Bassett, W.~Enck, and T.~Xie.
\newblock A study of grayware on google play.
\newblock In {\em Proceedings of the IEEE Security and Privacy Workshops},
  2016.

\bibitem{appbrain}
{Monetize, advertise and analyze Android apps}, 2017.
\newblock \url{www.appbrain.com/}.

\bibitem{arp2014drebin}
D.~Arp, M.~Spreitzenbarth, H.~Gascon, K.~Rieck, and C.~Siemens.
\newblock Drebin: Effective and explainable detection of android malware in
  your pocket.
\newblock In {\em Proceedings of the Network and Distributed System Security
  Symposium (NDSS)}, 2014.

\bibitem{pscout-paper}
K.~W.~Y. Au, Y.~F. Zhou, Z.~Huang, and D.~Lie.
\newblock Pscout: analyzing the android permission specification.
\newblock In {\em Proceedings of the ACM SIGSAC conference on Computer and
  communications security (CCS)}, 2012.

\bibitem{avdiienko2017detecting}
V.~Avdiienko, K.~Kuznetsov, I.~Rommelfanger, A.~Rau, A.~Gorla, and A.~Zeller.
\newblock Detecting behavior anomalies in graphical user interfaces.
\newblock In {\em Proceedings of the International Conference on Software
  Engineering Companion (ICSE-C)}, 2017.

\bibitem{ReliableLibrary}
M.~Backes, S.~Bugiel, and E.~Derr.
\newblock Reliable third-party library detection in android and its security
  applications.
\newblock In {\em Proceedings of the ACM SIGSAC Conference on Computer and
  Communications Security (CCS)}, 2016.

\bibitem{bartel2012automatically}
A.~Bartel, J.~Klein, Y.~Le~Traon, and M.~Monperrus.
\newblock Automatically securing permission-based software by reducing the
  attack surface: An application to android.
\newblock In {\em Proceedings of the IEEE/ACM International Conference on
  Automated Software Engineering (ASE)}, 2012.

\bibitem{bierma2014andlantis}
M.~Bierma, E.~Gustafson, J.~Erickson, D.~Fritz, and Y.~R. Choe.
\newblock Andlantis: Large-scale android dynamic analysis.
\newblock {\em arXiv preprint arXiv:1410.7751}, 2014.

\bibitem{block}
{How to Access Google Play Store in China?}, 2017.
\newblock
  \url{https://www.bestvpnprovider.com/how-to-access-google-play-store-china/}.

\bibitem{bohmer2011falling}
M.~B{\"o}hmer, B.~Hecht, J.~Sch{\"o}ning, A.~Kr{\"u}ger, and G.~Bauer.
\newblock Falling asleep with angry birds, facebook and kindle: a large scale
  study on mobile application usage.
\newblock In {\em Proceedings of the International conference on Human computer
  interaction with mobile devices and services}, 2011.

\bibitem{calciati2017apps}
P.~Calciati and A.~Gorla.
\newblock How do apps evolve in their permission requests?: a preliminary
  study.
\newblock In {\em Proceedings of the International Conference on Mining
  Software Repositories (MSR)}, 2017.

\bibitem{Longitudinal}
B.~Carbunar and R.~Potharaju.
\newblock A longitudinal study of the google app market.
\newblock In {\em Proceedings of the IEEE/ACM International Conference on
  Advances in Social Networks Analysis and Mining}, 2015.

\bibitem{chia2012app}
P.~H. Chia, Y.~Yamamoto, and N.~Asokan.
\newblock Is this app safe?: a large scale study on application permissions and
  risk signals.
\newblock In {\em Proceedings of the International conference on World Wide Web
  (WWW)}, 2012.

\bibitem{DNADroid}
J.~Crussell, C.~Gibler, and H.~Chen.
\newblock Attack of the clones: detecting cloned applications on {Android}
  markets.
\newblock In {\em Proceedings of the European Symposium on Research in Computer
  Security (ESORICS)}, 2012.

\bibitem{AdDarwin}
J.~Crussell, C.~Gibler, and H.~Chen.
\newblock Scalable semantics-based detection of similar {Android} applications.
\newblock In {\em Proceedings of the European Symposium on Research in Computer
  Security (ESORICS)}, 2013.

\bibitem{Dong-FSE-18}
F.~Dong, H.~Wang, L.~Li, Y.~Guo, T.~F. Bissyandé, T.~Liu, G.~Xu, and J.~Klein.
\newblock Frauddroid: Automated ad fraud detection for android apps.
\newblock In {\em Proceedings of the ACM Joint European Software Engineering
  Conference and Symposium on the Foundations of Software Engineering
  (ESEC/FSE)}, 2018.

\bibitem{stowaway}
A.~P. Felt, E.~Chin, S.~Hanna, D.~Song, and D.~Wagner.
\newblock Android permissions demystified.
\newblock In {\em Proceedings of the ACM conference on Computer and
  communications security (CCS)}, 2011.

\bibitem{gibler2012androidleaks}
C.~Gibler, J.~Crussell, J.~Erickson, and H.~Chen.
\newblock Androidleaks: Automatically detecting potential privacy leaks in
  android applications on a large scale.
\newblock In {\em Proceedings of the 5th international conference on Trust and
  Trustworthy Computing (TRUST)}, 2012.

\bibitem{AdRob}
C.~Gibler, R.~Stevens, J.~Crussell, H.~Chen, H.~Zang, and H.~Choi.
\newblock {AdRob}: examining the landscape and impact of {Android} application
  plagiarism.
\newblock In {\em Proceedings of the International Conference on Mobile
  Systems, Applications, and Services (MobiSys)}, 2013.

\bibitem{RiskRanker}
M.~Grace, Y.~Zhou, Q.~Zhang, S.~Zou, and X.~Jiang.
\newblock Riskranker: Scalable and accurate zero-day android malware detection.
\newblock In {\em Proceedings of the International Conference on Mobile
  Systems, Applications, and Services (MobiSys)}, 2012.

\bibitem{Juxtapp}
S.~Hanna, L.~Huang, E.~Wu, S.~Li, C.~Chen, and D.~Song.
\newblock Juxtapp: a scalable system for detecting code reuse among {Android}
  applications.
\newblock In {\em Proceedings of the International Conference on Detection of
  Intrusions and Malware and Vulnerability Assessment (DIMVA)}, 2012.

\bibitem{hassan2017empirical}
S.~Hassan, W.~Shang, and A.~E. Hassan.
\newblock An empirical study of emergency updates for top android mobile apps.
\newblock {\em Empirical Software Engineering}, 22(1):505--546, 2017.

\bibitem{hu2018dating}
Y.~Hu, H.~Wang, Y.~Zhou, Y.~Guo, L.~Li, B.~Luo, and F.~Xu.
\newblock Dating with scambots: Understanding the ecosystem of fraudulent
  dating applications.
\newblock {\em arXiv preprint arXiv:1807.04901}, 2018.

\bibitem{ikram2016analysis}
M.~Ikram, N.~Vallina-Rodriguez, S.~Seneviratne, M.~A. Kaafar, and V.~Paxson.
\newblock An analysis of the privacy and security risks of android vpn
  permission-enabled apps.
\newblock In {\em Proceedings of the Internet Measurement Conference (IMC)},
  2016.

\bibitem{ishii2017understanding}
Y.~Ishii, T.~Watanabe, F.~Kanei, Y.~Takata, E.~Shioji, M.~Akiyama, T.~Yagi,
  B.~Sun, and T.~Mori.
\newblock Understanding the security management of global third-party android
  marketplaces.
\newblock In {\em Proceedings of the ACM SIGSOFT International Workshop on App
  Market Analytics}, 2017.

\bibitem{Sumon}
S.~M. Kywe, Y.~Li, R.~H. Deng, and J.~Hong.
\newblock Detecting camouflaged applications on mobile application markets.
\newblock In {\em Proceedings of the International Conference on Information
  Security and Cryptology}, 2014.

\bibitem{li2015iccta}
L.~Li, A.~Bartel, T.~F. Bissyand{\'e}, J.~Klein, Y.~Le~Traon, S.~Arzt,
  S.~Rasthofer, E.~Bodden, D.~Octeau, and P.~Mcdaniel.
\newblock {IccTA: Detecting Inter-Component Privacy Leaks in Android Apps}.
\newblock In {\em Proceedings of the International Conference on Software
  Engineering (ICSE)}, 2015.

\bibitem{li2016investigation}
L.~Li, T.~F. Bissyand{\'e}, J.~Klein, and Y.~Le~Traon.
\newblock An investigation into the use of common libraries in android apps.
\newblock In {\em Proceedings of the IEEE International Conference on Software
  Analysis, Evolution, and Reengineering (SANER)}, 2016.

\bibitem{li2018cid}
L.~Li, T.~F. Bissyand{\'e}, H.~Wang, and J.~Klein.
\newblock Cid: automating the detection of api-related compatibility issues in
  android apps.
\newblock In {\em Proceedings of the ACM SIGSOFT International Symposium on
  Software Testing and Analysis (ISSTA)}, 2018.

\bibitem{li2017androzoo}
L.~Li, J.~Gao, M.~Hurier, P.~Kong, T.~F. Bissyand{\'e}, A.~Bartel, J.~Klein,
  and Y.~Le~Traon.
\newblock Androzoo++: Collecting millions of android apps and their metadata
  for the research community.
\newblock {\em arXiv preprint 1709.05281}, 2017.

\bibitem{li2017fbs}
Z.~Li, W.~Wang, C.~Wilson, J.~Chen, C.~Qian, T.~Jung, L.~Zhang, K.~Liu, X.~Li,
  and Y.~Liu.
\newblock Fbs-radar: Uncovering fake base stations at scale in the wild.
\newblock In {\em Proceedings of the Network and Distributed System Security
  Symposium (NDSS)}, 2017.

\bibitem{li2016exploring}
Z.~Li, W.~Wang, T.~Xu, X.~Zhong, X.-Y. Li, Y.~Liu, C.~Wilson, and B.~Y. Zhao.
\newblock Exploring cross-application cellular traffic optimization with baidu
  trafficguard.
\newblock In {\em Proceedings of the USENIX Symposium on Networked Systems
  Design and Implementation (NSDI)}, 2016.

\bibitem{libradargit}
{LibRadar - A detecting tool for 3rd-party libraries in Android apps}, 2017.
\newblock \url{https://github.com/pkumza/LibRadar}.

\bibitem{PEDAL}
B.~Liu, B.~Liu, H.~Jin, and R.~Govindan.
\newblock Efficient privilege de-escalation for ad libraries in mobile apps.
\newblock In {\em Proceedings of the International Conference on Mobile
  Systems, Applications, and Services (MobiSys)}, 2015.

\bibitem{liu2016identifying}
M.~Liu, H.~Wang, Y.~Guo, and J.~Hong.
\newblock Identifying and analyzing the privacy of apps for kids.
\newblock In {\em Proceedings of the International Workshop on Mobile Computing
  Systems and Applications (HotMobile)}, 2016.

\bibitem{Censorship}
Z.~Lu, Z.~Li, J.~Yang, T.~Xu, E.~Zhai, Y.~Liu, and C.~Wilson.
\newblock Accessing google scholar under extreme internet censorship: A legal
  avenue.
\newblock In {\em Proceedings of the ACM/IFIP/USENIX Middleware Conference:
  Industrial Track (Middleware)}, 2017.

\bibitem{libradar}
Z.~Ma, H.~Wang, Y.~Guo, and X.~Chen.
\newblock Libradar: Fast and accurate detection of third-party libraries in
  android apps.
\newblock In {\em Proceedings of the International Conference on Software
  Engineering Companion (ICSE-C)}, 2016.

\bibitem{ADDetect}
A.~Narayanan, L.~Chen, and C.~K. Chan.
\newblock Addetect: Automated detection of android ad libraries using semantic
  analysis.
\newblock In {\em Proceedings of the IEEE International Conference on
  Intelligent Sensors, Sensor Networks and Information Processing (ISSNIP)},
  2014.

\bibitem{ng2014android}
Y.~Y. Ng, H.~Zhou, Z.~Ji, H.~Luo, and Y.~Dong.
\newblock Which android app store can be trusted in china?
\newblock In {\em Proceedings of the IEEE Computer Society International
  Conference on Computers, Software and Applications (COMPSAC)}, 2014.

\bibitem{MeasurementEcosystem}
T.~Petsas, A.~Papadogiannakis, M.~Polychronakis, E.~P. Markatos, and
  T.~Karagiannis.
\newblock Measurement, modeling, and analysis of the mobile app ecosystem.
\newblock {\em ACM Trans. Model. Perform. Eval. Comput. Syst.}, 2(2):7:1--7:33,
  Mar. 2017.

\bibitem{privacygrade}
{Privacy Grade}, 2017.
\newblock \url{www.privacygrade.org}.

\bibitem{razaghpanah2018apps}
A.~Razaghpanah, R.~Nithyanand, N.~Vallina-Rodriguez, S.~Sundaresan, M.~Allman,
  C.~Kreibich, and P.~Gill.
\newblock Apps, trackers, privacy, and regulators: A global study of the mobile
  tracking ecosystem.
\newblock In {\em Proceedings of the Network and Distributed System Security
  Symposium (NDSS)}, 2018.

\bibitem{razaghpanah2015haystack}
A.~Razaghpanah, N.~Vallina-Rodriguez, S.~Sundaresan, C.~Kreibich, P.~Gill,
  M.~Allman, and V.~Paxson.
\newblock Haystack: In situ mobile traffic analysis in user space.
\newblock {\em ArXiv e-prints}, 2015.

\bibitem{ren2018bug}
J.~Ren, M.~Lindorfer, D.~J. Dubois, A.~Rao, D.~Choffnes, and
  N.~Vallina-Rodriguez.
\newblock Bug fixes, improvements,... and privacy leaks.
\newblock In {\em Proceedings of the Network and Distributed System Security
  Symposium (NDSS)}, 2018.

\bibitem{ren2016recon}
J.~Ren, A.~Rao, M.~Lindorfer, A.~Legout, and D.~Choffnes.
\newblock Recon: Revealing and controlling pii leaks in mobile network traffic.
\newblock In {\em Proceedings of the International Conference on Mobile
  Systems, Applications, and Services (MobiSys)}, 2016.

\bibitem{AVclass}
M.~Sebasti{\'a}n, R.~Rivera, P.~Kotzias, and J.~Caballero.
\newblock Avclass: A tool for massive malware labeling.
\newblock In {\em Proceedings of the International Symposium on Research in
  Attacks, Intrusions, and Defenses (RAID)}, pages 230--253. Springer, 2016.

\bibitem{smartphone}
{List of countries by smartphone penetration}, 2017.

\bibitem{sounthiraraj2014smv}
D.~Sounthiraraj, J.~Sahs, G.~Greenwood, Z.~Lin, and L.~Khan.
\newblock Smv-hunter: Large scale, automated detection of ssl/tls
  man-in-the-middle vulnerabilities in android apps.
\newblock In {\em Proceedings of the Network and Distributed System Security
  Symposium (NDSS)}, 2014.

\bibitem{vallina2012breaking}
N.~Vallina-Rodriguez, J.~Shah, A.~Finamore, Y.~Grunenberger, K.~Papagiannaki,
  H.~Haddadi, and J.~Crowcroft.
\newblock {Breaking for Commercials: Characterizing Mobile Advertising}.
\newblock In {\em Proceedings of the ACM Internet Measurement Conference
  (IMC)}, 2012.

\bibitem{MeasurementGoogle}
N.~Viennot, E.~Garcia, and J.~Nieh.
\newblock A measurement study of google play.
\newblock In {\em Proceedings of the International Conference on Measurement
  and Modeling of Computer Systems (SIGMETRICS)}, 2014.

\bibitem{wang2017understanding}
H.~Wang and Y.~Guo.
\newblock Understanding third-party libraries in mobile app analysis.
\newblock In {\em Proceedings of the IEEE/ACM International Conference on
  Software Engineering Companion (ICSE-C)}, 2017.

\bibitem{wukong}
H.~Wang, Y.~Guo, Z.~Ma, and X.~Chen.
\newblock Wukong: A scalable and accurate two-phase approach to android app
  clone detection.
\newblock In {\em Proceedings of the International Symposium on Software
  Testing and Analysis (ISSTA)}, 2015.

\bibitem{wang2015reevaluating}
H.~Wang, Y.~Guo, Z.~Tang, G.~Bai, and X.~Chen.
\newblock Reevaluating android permission gaps with static and dynamic
  analysis.
\newblock In {\em Proceedings of the IEEE Global Communications Conference
  (GLOBECOM)}, 2015.

\bibitem{wang2015using}
H.~Wang, J.~Hong, and Y.~Guo.
\newblock Using text mining to infer the purpose of permission use in mobile
  apps.
\newblock In {\em Proceedings of the ACM International Joint Conference on
  Pervasive and Ubiquitous Computing (UbiComp)}, 2015.

\bibitem{WangRemoval}
H.~Wang, H.~Li, L.~Li, Y.~Guo, and G.~Xu.
\newblock {Why are Android Apps Removed From Google Play? A Large-scale
  Empirical Study}.
\newblock In {\em Proceedings of the International Conference on Mining
  Software Repositories (MSR)}, 2018.

\bibitem{wang2017understandingPurpose}
H.~Wang, Y.~Li, Y.~Guo, Y.~Agarwal, and J.~I. Hong.
\newblock Understanding the purpose of permission use in mobile apps.
\newblock {\em ACM Transactions on Information Systems (TOIS)}, 35(4):43, 2017.

\bibitem{Ecosystem}
H.~Wang, Z.~Liu, Y.~Guo, X.~Chen, M.~Zhang, G.~Xu, and J.~Hong.
\newblock An explorative study of the mobile app ecosystem from app developers'
  perspective.
\newblock In {\em Proceedings of the International Conference on World Wide Web
  (WWW)}, 2017.

\bibitem{wang2016measuring}
T.~Wang, D.~Wu, J.~Zhang, M.~Chen, and Y.~Zhou.
\newblock Measuring and analyzing third-party mobile game app stores in china.
\newblock {\em IEEE Transactions on Network and Service Management},
  13(4):793--805, 2016.

\bibitem{wei2017deep}
F.~Wei, Y.~Li, S.~Roy, X.~Ou, and W.~Zhou.
\newblock Deep ground truth analysis of current android malware.
\newblock In {\em Proceedings of the International Conference on Detection of
  Intrusions and Malware, and Vulnerability Assessment (DIMVA)}, 2017.

\bibitem{wu2013impact}
L.~Wu, M.~Grace, Y.~Zhou, C.~Wu, and X.~Jiang.
\newblock The impact of vendor customizations on android security.
\newblock In {\em Proceedings of the ACM SIGSAC Conference on Computer and
  communications security (CCS)}, 2013.

\bibitem{yang2017characterizing}
X.~Yang, D.~Lo, L.~Li, X.~Xia, T.~F. Bissyand{\'e}, and J.~Klein.
\newblock Characterizing malicious android apps by mining topic-specific data
  flow signatures.
\newblock {\em Information and Software Technology}, 2017.

\bibitem{viewdroid}
F.~Zhang, H.~Huang, S.~Zhu, D.~Wu, and P.~Liu.
\newblock {ViewDroid}: towards obfuscation-resilient mobile application
  repackaging detection.
\newblock In {\em Proceedings of the ACM Conference on Security and Privacy in
  Wireless and Mobile Networks (WiSec)}, 2014.

\bibitem{zheng2012adam}
M.~Zheng, P.~P. Lee, and J.~C. Lui.
\newblock Adam: an automatic and extensible platform to stress test android
  anti-virus systems.
\newblock In {\em Proceedings of the International conference on Detection of
  Intrusions and Malware, and Vulnerability assessment (DIMVA)}, 2012.

\bibitem{Piggyback}
W.~Zhou, Y.~Zhou, M.~Grace, X.~Jiang, and S.~Zou.
\newblock Fast, scalable detection of ``piggybacked'' mobile applications.
\newblock In {\em Proceedings of the ACM Conference on Data and Application
  Security and Privacy (CODASPY)}, 2013.

\bibitem{DroidMoss}
W.~Zhou, Y.~Zhou, X.~Jiang, and P.~Ning.
\newblock Detecting repackaged smartphone applications in third-party {Android}
  marketplaces.
\newblock In {\em Proceedings of the ACM Conference on Data and Application
  Security and Privacy (CODASPY)}, 2012.

\bibitem{zhou2012dissecting}
Y.~Zhou and X.~Jiang.
\newblock Dissecting android malware: Characterization and evolution.
\newblock In {\em Proceedings of the IEEE Symposium on Security and Privacy
  (S\&P)}, 2012.

\end{thebibliography}

\end{document}